\documentclass[lettersize,journal]{IEEEtran}
\usepackage{amsmath,amsfonts}
\usepackage{algorithmic}
\usepackage{algorithm}
\usepackage{array}
\usepackage[caption=false,font=normalsize,labelfont=sf,textfont=sf]{subfig}
\usepackage{textcomp}
\usepackage{stfloats}
\usepackage{url}
\usepackage{verbatim}
\usepackage{graphicx}
\usepackage{cite}
\usepackage{booktabs}
\usepackage{color}
\usepackage{subfig}
\usepackage{setspace} 
\usepackage{newtxtext,newtxmath}

\begin{document}

\title{QoE-oriented Dependent Task Scheduling under Multi-dimensional QoS Constraints over Distributed Networks}

\author{Xuwei Fan,~\IEEEmembership{Student Member,~IEEE,} Zhipeng Cheng,~\IEEEmembership{Member,~IEEE,} Ning  Chen,~\IEEEmembership{Student Member,~IEEE,} \\ Lianfen Huang,~\IEEEmembership{Member,~IEEE,} and Xianbin Wang,~\IEEEmembership{Fellow,~IEEE}
\thanks{Xuwei Fan, Ning Chen and Lianfen Huang are with the Department of Information and Communication Engineering, Xiamen University, Xiamen 361005, China (e-mail: xwfan@stu.xmu.edu.cn; ningchen@stu.xmu.edu.cn; lfhuang@xmu.edu.cn)}
\thanks{Zhipeng Cheng is with the School of Future Science and Engineering, Soochow University, Suzhou 215006, China (e-mail: chengzp\_x@163.com).}
\thanks{Xianbin Wang is with the Department of Electrical and Computer Engineering, Western University, London, ON N6A 5B9, Canada (e-mail: xianbin.wang@uwo.ca).}}

\markboth{}%
{Shell \MakeLowercase{\textit{et al.}}: A Sample Article Using IEEEtran.cls for IEEE Journals}


\maketitle

\begin{abstract}
Task scheduling as an effective strategy can improve application performance on computing resource-limited devices over distributed networks. However, existing evaluation mechanisms fail to depict the complexity of diverse applications, which involve dependencies among tasks, computing resource requirements, and multi-dimensional quality of service (QoS) constraints. Furthermore, traditional QoS-oriented task scheduling strategies struggle to meet the performance requirements without considering differences in satisfaction and acceptance of application, leading application failures and resource wastage. To tackle these issues, a quality of experience (QoE) cost model is designed to evaluate application completion, depicting the relationship among application satisfaction, communications, and computing resources in the distributed networks. Specifically, considering the sensitivity and preference of QoS, we model the different dimensional QoS degradation cost functions for dependent tasks, which are then integrated into the QoE cost model. Based on the QoE model, the dependent task scheduling problem is formulated as the minimization of overall QoE cost, aiming to improve the application performance in the distributed networks, which is proven Np-hard. Moreover, a heuristic Hierarchical Multi-queue Task Scheduling Algorithm (HMTSA) is proposed to address the QoE-oriented task scheduling problem among multiple dependent tasks, which utilizes hierarchical multiple queues to determine the optimal task execution order and location according to different dimensional QoS priorities. Finally, extensive experiments demonstrate that the proposed algorithm can significantly improve the satisfaction of applications.
\end{abstract}


\section{Introduction}
\IEEEPARstart{W}{ith} the unprecedented evolution of communication and computing technologies, more connected devices, smart applications, and massive amount of data have surged explosively. According to IDC's predictions, there will be 55 billion connected Internet of Technology (IoT) devices worldwide by 2025, which support a wide variety of computing-intensive applications, e.g., autonomous driving, virtual reality, and medical diagnosis. Most of these applications are intelligent and diversified, comprising dependent tasks \cite{ref16}.  For instance, a medical diagnosis application used in telemedicine involves a series of dependent tasks, including data acquisition, preprocessing, image analysis, diagnosis, and report generation. Each of these dependent tasks relies on the output of the preceding tasks and processes according to the logical order. However, these increasing complexity and service requirements of applications bring challenges for the lightweight IoT devices, originally design for better portability and usability. To address this challenge, task scheduling as an effective strategy allows computational tasks to be scheduled across different devices for ultilizing distributed computing resources, thereby reducing execution latency and providing satisfactory performance of applications through cooperative processing.  

\subsection{Motivation} 
Empowered by the evolution of information and communication technologies, a new era of user-centric diverse service provisioning has arrived, which generates numerous smart computing applications and services (e.g., auxiliary diagnosis, decision making, and data analysis). However, existing wireless infrastructure, e.g., 5G, mainly focus on supporting quality of service (QoS) in communication scenarios, leading to the new challenges for satisfying such `beyond communication' services \cite{ref2}. With the rapid convergence of communications and computing, future wireless communication systems and networks including B5G and future 6G are expected to provide application tailored QoS by considering computing capacity, supporting a wide range of computing applications with stringent QoS requirements in terms of latency, reliability, energy efficiency \cite{ref1,ref3}. 

In order to meet these QoS requirements leveraging distributed computation capabilities, task scheduling and offloading have been proposed as meaningful strategies based on QoS optimization objective \cite{ref4,ref6,ref7}. However, existing QoS evaluation mechanisms face limitations in characterizing the diverse requirements of heterogeneous services and applications without considering dimensions and sensitivities of QoS requirements of diverse applications, which leads to the ineffectiveness of scheduling strategies. Specifically, many emerging applications, particularly those related to machine communications have brought additional dimensions of QoS requirements in terms of latency and reliability \cite{ref14,ref15}. However, most works aim at static single dimensional QoS guarantee or QoS maximization (e.g., latency and energy efficiency), leading to resource wastage by blindly pursuing application-agnostic QoS targets. Moreover, the differences in satisfaction and acceptance of diverse applications contribute to the varying sensitivities of QoS, which have been ignored in this existing studies, posing a critical challenge for scheduling tasks from diverse applications. For example, applications could have very different levels of sensitivity to application completion deadlines (i.e., hard deadline threshold or soft deadline threshold). In the case of a hard-deadline based application, the application has to be processed within a specified deadline; otherwise, the application will become useless. On the other hand, a soft-deadline application is valid up to some extent with performance degradation for exceeding the deadline. 

Quality of experience (QoE) widely perceived as an effective application-sepcific performance indicator, focuses on users' satisfaction when using a particular application or service over a network \cite{ref8,ref9}, which can depict the influence by dynamic QoS provisioning of the network (e.g., transmission rate and bit error rate) \cite{ref10,ref11}. Many literatures  have presented QoE models that jointly depend on the user preference and the QoS parameters of network, which have been proved in optimizing the utilization of resoures by supporting acceptable services over the pursuit of ultimate performance \cite{ref12,ref13}. However, in the case of `beyond communication' services, QoS constraints are related not only to QoS parameters of network, but also to the computing resource requirements. Moreover, the inherent structure of applications introduces further complexity into defining a QoE model that characterizes the cumulative impact of processing each task. Based on these, redefining a precise QoE model capable of capturing to diverse dimensions and sensitivities of QoS parameters while accurately reflecting the satisfaction of different applications is a crucial prerequisite for effectively utilizing the computing resources to schedule the dependent tasks. Nevertheless, it is still a challenge to integrate the factors involving  QoS constraints, resource requirements, and network conditions into the QoE model.

On the other hand, dependent task scheduling over distributed networks still faces many challenges to improve the performance of application due to the decentralization and heterogeneity of devices and dependencies of applications \cite{ref17, ref18}. Firstly, the abundance of heterogeneous devices and communication links in distributed networks raises the complexity of the design of dependent task scheduling algorithm. Existing scheduling algorithms strive for optimality through sophisticated designs in mobile computing networks, but this pursuit of optimality often introduces higher complexity, leading unacceptable algorithmic complexity in the distributed network \cite{ref22,ref23}. Balancing performance and complexity of scheduling algirithm poses a significant challenge in the context of distributed network. Secondly, communication between devices via multi-hop connections breaks down the physical isolation of computation, further complicating the design introducing the second challenge: how to map multiple dependent tasks to distributed nodes within multiple hops \cite{ref24,ref25}. Finally, the challenge is that multiple dependent tasks with different QoS dimensions and sensitivities demand scheduling algorithms that can intelligently schedule dependent tasks based on QoS constraints and communication-computation requirements, ultimately enhancing the overall QoE \cite{ref26,ref27}.

Although the above works have made significant contributions in dependent tasks scheduling and offloading, it is still a challenge to schedule dramatically increased variety of applications to satisfy their performance requirements under various dimensions and sensitivities of QoS constraints over distributed networks. In summary, we should resolve the following challenges to realize effective scheduling of dependent tasks under multi-dimensional QoS constraints:

Question 1: How to develop a QoE model that is capable of evaluating the completion of various dependent tasks under multi-dimensional QoS constraints?

Question 2: How to design an algorithm for QoE-oriented dependent task scheduling under multi-dimensional QoS constraints to maximize overall devices' QoE over distributed networks with acceptable computational complexity?

\subsection{ Contributions} 
To the best of our knowledge, this paper is among the first to study the QoE-oriented dependent task scheduling problem under multi-dimensional QoS constraints over distributed networks. Multiple applications and the dependencies among tasks introduce challenges in scheduling and offloading tasks to spare computing nodes. Moreover, the multi-dimensional QoS constraints and applications' preference pose the challenges in prioritizing tasks from multiple applications, that further complicates the algorithm design. Motivated by which, we propose an efficient  and heuristic Hierarchical Multi-queue Task Scheduling Algorithm (HMTSA) to prioritize dependent tasks and assign them into spare computing nodes. Major contributions are summarized below: 
\begin{itemize}
\item[1)] An interesting multiple dependent task scheduling problem is studied, where each application is modeled as a Directed Acyclic Graph (DAG) with task nodes and weighted edges with multi-dimensional QoS constraints. These constraints are categorized into two classes in terms of long term QoS parameter and short term QoS parameter depending on whether the QoS parameter is influenced  by the time and order of task placement. Moreover, a QoE cost model is developed to evaluate completion of each application under multi-dimensional QoS constraints. The objective of the multiple dependent task scheduling problem is minimizing the overall QoE cost of the applications with multi-dimensional QoS constraints.

\item[2)] The afore-mentioned problem is formulated as an integer non-linear programming problem, which is NP-hard. We propose an efficient QoE-oriented HMTSA to prioritize the tasks by multiple queues based on the dynamic updating task priority. Firstly, HMTSA determines the number of tasks assigning for an application based on the application-level priority and then determines the task assignment order according to the task-level priority. Through our algorithm, urgent applications can schedule more tasks in a single round of scheduling, while ensuring that tasks of those applications with high reliability requirements are assigned to nodes with better transmission channels.

\item[3)] Through comprehensive numerical analysis and comparative evaluations, the proposed HMTSA demonstrates superior performance compared to baseline methods across various problem sizes, particularly in scenarios with significantly limited computing resources.
\end{itemize}

The rest of this paper is organized as follows. Section \uppercase\expandafter{\romannumeral2} discusses the related works. Section \uppercase\expandafter{\romannumeral3} introduces the system model and problem formulation.  A  detailed algorithm description of HMTSA is provided in Section \uppercase\expandafter{\romannumeral4}. The numerical analysis and experimental simulation of our proposed algorithm are conducted in Section \uppercase\expandafter{\romannumeral5}. Finally, Section \uppercase\expandafter{\romannumeral6} concludes the paper.
\section{Related work}
The accurate evaluation of application or service completion is a prerequisite for effective task scheduling. In recent years, QoS and QoE, as typical indicators, have undergone extensive studies on the design of task scheduling strategies aimed at enhancing application performance and optimizing resource utilization. The related work can be categorized into QoS and QoE modeling, along with QoS-oriented and QoE-oriented dependent task scheduling, as follows.
\subsection{QoS and QoE modeling}
Existing QoS models are implicit in the optimization objective, primarily concentrating on QoS guaranteeing, QoS maximizing and striking a trade-off between latency and energy consumption \cite{ref4,ref6,ref7}. Some works strive to either maximize QoS, minimize QoS cost or find a balance among different QoS parameters within soft threshold QoS models. Cheng et al. introduced a method utilizing MADDPG for task partitioning and scheduling to minimize average execution delay as well as average energy consumption simultaneously \cite{ref6}. Qi et al. proposed a multi-task deep reinforcement learning approach for scalable parallel task scheduling, aiming to strike a balance between delay and energy consumption \cite{ref7}. Conversely, other works are dedicated to ensuring QoS within the hard threshold QoS model. Hu et al. delved into scheduling task to minimizes the system energy consumption under the constraints of response time \cite{r1}. Recently, a few works have been proposed for dependent tasks scheduling to shortening latency \cite{ref22,ref23}, saving energy \cite{ref24,ref25}, minimizing deadline violation ratio \cite{ref26,ref27}, as well as achieving the trade-off between delay and energy consumption \cite{ref28,ref29}. However, the continuous increase in resources may result in only marginal improvements in QoE, even with better QoS, once the QoE metrics surpass the perception limits of the application considering the multiple dimensions and sensitivities of QoS constraints. 

An accurate QoE evaluation model is a prerequisite for facilitating efficient resource allocation and task scheduling to improve the user experience with limited resources \cite{ref32}. Initially, QoE models were primarily employed in video applications (e.g., streaming video, VR and AR) to enhance user's experience \cite{ref33, ref43}. Recently, QoE model has gradually been expanded to other scenario, divided into two categories: mathematic based models and machine learning based models \cite{ref3, ref43}. Wang et al. categorized the influencing QoE factors into four aspects: human, technology, business, and context. Then they employed a deep learning method to model the relation between QoE and influencing factors \cite{ref3}. In \cite{ref43}, a personalized QoE improvement scheme fully leveraging the time-varying influences on users' QoE, including user-awareness, device-awareness and context-awareness was proposed and then an efficient deep learning based model was employed to precisely characterize personalization. Zabetian et al. present a machine learning algorithm to model the mean opinion score (MOS) of the call service in terms of the received signal strength indicator (RSSI) \cite{ref34}. These works aim to establish the accuracy of QoE and QoS but tend to overlook the complexity of applications, the heterogeneity of demands, and the multi-dimension of network systems. Therefore, establishing a QoE model that accurately reflects the complex network and application preferences remains a challenge.

\subsection{QoS-oriented and QoE-oriented dependent task scheduling}
QoS-oriented dependent task scheduling strategies have been explored extensively  \cite{ref22,ref23,ref24,ref25,ref26,ref27}. Li et al. tackled the scheduling of precedence-constrained tasks in a fog computing environment for a mobile application, aiming to strike a trade-off between energy and time using an alternate optimization algorithm \cite{r2}. Dai et al. utilized the multi-armed bandit theory to develop an online learning-based dependent task offloading algorithm, reducing application delay while adhering to long-term energy budget constraints in edge computing with unknown system-side information \cite{r3}. Xiao et al. proposed a two-stage dependent task scheduling algorithm to achieve low latency while controlling energy consumption. Initially, a priority scheduling strategy is designed to determine the task order. Subsequently, an offloading scheme based on reinforcement learning is introduced to assign tasks to computing nodes in the edge computing network \cite{ref41}. These QoS-oriented algorithms are tailored to efficiently schedule dependent tasks, improving application performance in edge computing networks. However, their effectiveness may degrade in more complex networks with multiple-hop transmissions. Moreover, some solutions necessitate time-consuming iterative searches for large-scale dependent tasks in distributed networks, which could be impractical for devising timely scheduling strategies, prompting the use of simple heuristic algorithms to optimize the scheduling problem. Lastly, these QoS-oriented algorithms fall short of effectively configuring resources and enhancing satisfaction without considering application preferences and multiple-dimensional QoS constraints \cite{ref30,ref31}.

QoE-oriented optimization schemes have been used in various scenarios \cite{ref35,ref36,ref42}. Lai et al. introduced a heuristic approach to allocate resources in support of latency-sensitive applications \cite{ref35}. In \cite{ref36}, an online deep reinforcement learning-based algorithm was proposed to orchestrate network functions and maximize overall QoE. Alquerm et al. developed a two-stage deep reinforcement learning scheme that effectively allocates edge resources to serve IoT applications and maximize users' QoE \cite{ref42}. However, these studies overlook the complexity of distributed networks and task dependencies, potentially causing the proposed algorithms to degrade in performance for dependent tasks under multi-dimensional and varying sensitive QoS constraints.

\begin{figure*}[!t]
\centering
\includegraphics[width=7in]{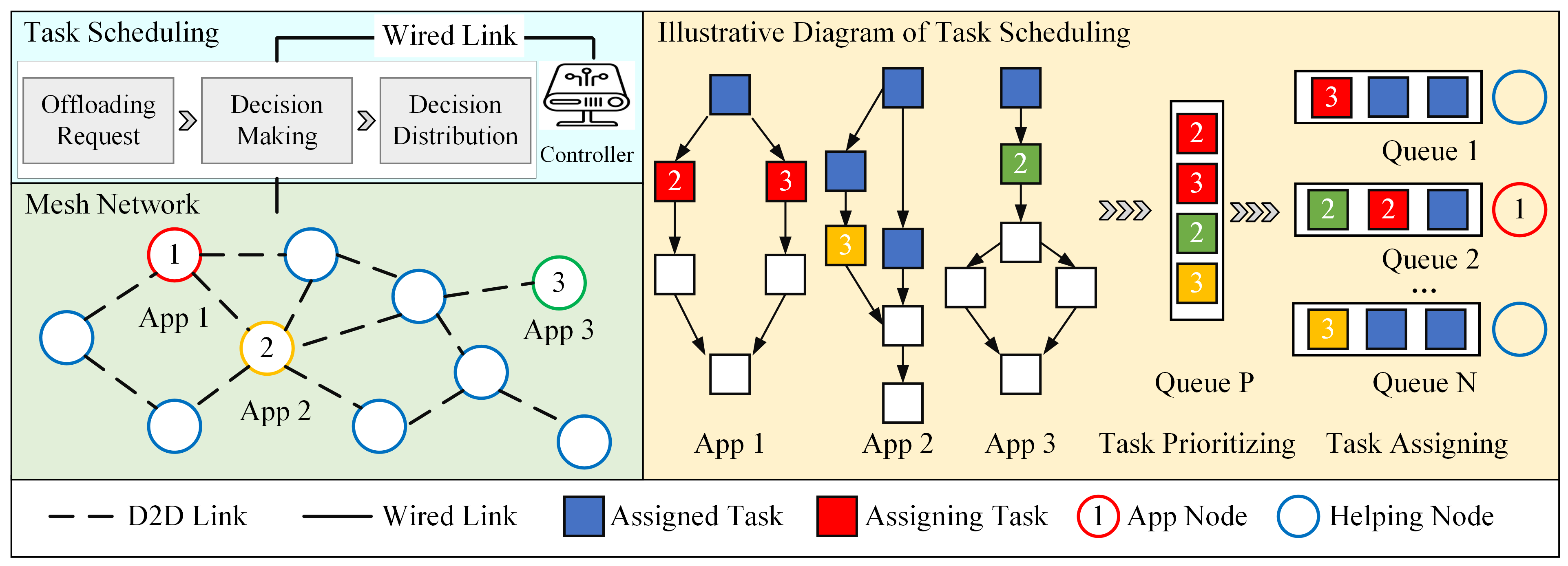}
\caption{The diagram of system model and dependent task scheduling in distributed network. }
\label{fig_1}
\end{figure*}
\section{System Model and Problem Formulation}
Without loss of generality, we consider a mesh network as the typical distributed network, in which each node can either be an IoT device or an edge computing device endowed with computing resources \cite{ref38}. As illustrated in Fig. 1,  some nodes, referred to as App Nodes, request services for application processing. These App Nodes generate applications consisting of dependent tasks with diverse resource requirements, QoS constraints, and preferences of QoS. Additionally, other nodes are available to assist the App Nodes by providing computing resources, regarded as Helping Nodes. Moreover, all nodes in the network establish links using device-to-device (D2D) communication.

The controller has the capability to capture the statue information reported by nodes (e.g., channel state information, utilization of computing resources, and application requirements). When App Nodes request tasks offloading, the controller orchestrates the tasks considering different QoS constraints, ensuring overall satisfactory QoE. Then, the controller generates a feasible task offloading decision, which is dispatched to the Helping Nodes. Finally, the tasks are offloaded and processed in assigned nodes according to the decision made by the controller.

In the task scheduling stage, tasks are prioritized in a hierarchical manner based on QoS degradation functions. Then, task assigning is performed to the optimal node according to the QoS constraints of the tasks. This scheduling strategy aims to effectively utilize resources and maximize the overall system's QoE. In the following, a detailed description of the system model is introduced. Note that TABLE \uppercase\expandafter{\romannumeral1} lists the frequently used notations.
\begin{table}[ht]
    \centering
    \caption{Frequently Used Notations}
    \label{tab:notation}
    \begin{tabular}{l|l}
        \toprule
        \textbf{Notation} & \textbf{Description} \\
        \midrule
        \footnotesize$N$ & \footnotesize The total number of nodes in the network\\
      \footnotesize $\mathcal{V}$ &\footnotesize The set of computing nodes in the network\\
	\footnotesize $\mathcal{E}$ &\footnotesize The set of transmission links between nodes \\
      \footnotesize  ${\overset{\sim}{\mathcal{V}}}_{n}(t)$ &\footnotesize The task set of app ${\overset{\sim}{\mathcal{G}}}_{n}(t)$\\
	\footnotesize${\overset{\sim}{\mathcal{E}}}_{n}(t)$  &\footnotesize The set of workload transmission of app ${\overset{\sim}{\mathcal{G}}}_{n}(t)$\\
	\footnotesize$v_n(t)$ &\footnotesize The computing node\\ 
	\footnotesize$v'_n(t)$ &\footnotesize The indicator for $v_n(t)$ requesting app processing\\ 
	\footnotesize$\overset{\sim}v_{n,j}(t)$ &\footnotesize The $j$-th task of task set $\overset{\sim}{\mathcal{V}}_{n}(t)$\\ 
	\footnotesize${\overset{\sim}{e}}_{i,i'}^{n}(t)$ &\footnotesize The workload transmission between task $\overset{\sim}v_{n,i}(t)$ \\
	& \footnotesize and $ \overset{\sim}v_{n,i'}(t)$\\
      \footnotesize $\mathbf{F}_m(t)$ &\footnotesize The computing resource of node $v_m(t)$ \\
	\footnotesize  $\mathbf{q}_{n,j}(t)$ &\footnotesize The indicator for the type of task $\overset{\sim}v_{n,j}(t)$ \\
      \footnotesize $\mathbf{r}_{n,j}(t)$ &\footnotesize The resource requirement of task $\overset{\sim}v_{n,j}(t)$\\
	\footnotesize $\mathbf{X}_j^n(t)$ &\footnotesize Vector indicating assignment strategy of $\overset{\sim}v_{n,j}(t)$ \\
	\footnotesize$\mathbf{V}'(t)$&\footnotesize Vector indicating if nodes request app processing  \\
	\footnotesize$d_n(t)$, $e_n(t)$&\footnotesize The deadline and the upper limit of data error \\
	&\footnotesize of ${\overset{\sim}{\mathcal{G}}}_{n}(t)$\\
	\footnotesize$w_{n}^d(t)$, $w_{n}^e(t)$&\footnotesize The preference of deadline and accuracy of  ${\overset{\sim}{\mathcal{G}}}_{n}(t)$\\
	\footnotesize$C_{j,j'}^{n}(t)$,$E_{j,j'}^{n}(t)$ & \footnotesize The data transmission time and data error rate \\
	&\footnotesize between task $\overset{\sim}v_{n,j}(t)$ and $\overset{\sim}v_{n,j'}(t)$\\
	\footnotesize$R_j^n(t)$, $T_j^n(t)$&\footnotesize The data error rate and execution time of  $\overset{\sim}v_{n,j}(t)$\\
	\footnotesize$R_j^{'n}(t)$, $T_j^{'n}(t)$&\footnotesize The estimate data error rate and execution time of\\
	&\footnotesize task $\overset{\sim}v_{n,j}(t)$\\

        \bottomrule
    \end{tabular}
\end{table}

\subsection{Model of Distributed Networks}
In the distributed mesh networks, devices and the connection of devices can be represented by a graph $\mathcal{G} = (\mathcal{V},\mathcal{E})$, where $\mathcal{V} = \left\{ v_{n} \middle| n \in \left\{ 1,\cdots,n,\cdots,N \right\} \right\}$ is the set of device with computing capacity and $N=|\mathcal{V}|$ indicates the number of devices in the distrubuted network. Each device $v_n$ has various types of computing resources denoted by vector $\mathbf{F}_n=\{{F_{n}^{cpu},F_{n}^{gpu},O_{n}}\}$, in which $F_n^{cpu}$, $F_n^{gpu}$ are frequency of CPU and GPU, respectively, and $O_n$ is I/O rate. The edge set $\mathcal{E} = \left\{ E_{m,m{'}} \middle| m,m{'} \in \left\{ {1,\cdots,n,\cdots,N} \right\},m \neq m' \right\}$ represents network links between node $v_m$ and node $v_{m'}$, which remains relatively stable throughout the application request processing. Assume that the transmission link $E_{m,m'}(t)$ is composed of $\{c_{m,m'} (t),e_{m,m'} (t)\}$ in time period $t$, where $c_{m,m'}(t)$ is the transmission rate from the device $v_m$ to device $v_{m'}$ and $e_{m,m'} (t)$ is bit error rate (BER)  of the link, respectively \footnote [1] {This paper mainly addresses the dependent task scheduling under multi-dimensional QoS constraints in the mesh network. To maintain the clarity of our approach, we assume that the entire network state information can be collected by the controller, and the transmission rate can be set as constants. This simplification can be readily extended to real channel models.}. For analytical simplicity, we neglect the band width constraints, and assume that the communication link $E_{m,m{'}}(t)=E_{m{'},m}(t)$.
Additionally, the task execution on each node follows First Come First Serve principle (FCFS) \footnote [2] {Multiple processors can be regarded as multiple devices with the same resources. Therefore, this assumption can be easily extended to multiple processors with FCFS principle. For the sake of readability, we consider single processor.}.

In real physical world, application processing request will arrive in arbitrary time. Therefore, we consider the application processing request in time period $t$, and task scheduling algorithm can be extended to other period. $\mathbf{V}{'}(t) = \left\{ v_{1}'(t),\cdots,v_{n}'(t),\cdots,v_{N}'(t) \right\}$ is a binary vector indicating if the device requests an application processing. If the application is generated by $v_n$, $v_n'(t)=1$, denoted as App Node; otherwise, $v_n'(t)=0$, denoted as Helping Node. In order to improve users' QoE, App Nodes offload tasks to Helping Nodes for parallel computation.  

\subsection{Model of Dependent Tasks}
In order to clarify the logical sequence of dependent tasks to design effective scheduling scheme, Directed Acyclic Graph (DAG) model was proposed and widely adapted to describe the dependencies between tasks \cite{ref17, ref18}. Applications are transferred into DAG based on their inherent attributes, where the task can be activated after the precursor of the dependent task is completed. Specifically, in time period $t$, each application is represented by a graph ${\overset{\sim}{\mathcal{G}}}_{n}(t) = \left( {\overset{\sim}{\mathcal{V}}}_{n}(t),{\overset{\sim}{\mathcal{E}}}_{n}(t) \right)$, where ${\overset{\sim}{\mathcal{V}}}_{n}(t) = \left\{ {\overset{\sim}{v}}_{n,j}(t) \middle| j \in \left\{ 1,\cdots,j,\cdots,J_{n} \right\} \right\}$ denotes the task set of application ${\overset{\sim}{\mathcal{G}}}_{n}(t)$, and $J_{n}$ indicates the number of tasks. ${\overset{\sim}{\mathcal{E}}}_{n}(t) = \left\{ {\overset{\sim}{e}}_{i,i'}^{n}(t) \middle| i,i^{'} \in \left\{ {1,\cdots,j,\cdots,J_{n}} \right\},i \neq i' \right\}$ denotes the set of workload for transmission. Such tasks can be scheduled into different nodes to accelerate the computing process. 

There are diverse QoS constrains of depende tasks in the real-life (e.g. latency, accuracy and energy efficiency), which can be classified into long term QoS parameter and short term QoS parameter as follows. 

\textbf{Defination 1.} (Long term QoS parameter) The long term QoS parameter is influenced throughout the entire time span of task scheduling, contingent upon the order and placement of task processing. For example, varying sequences of task assignment and processing may result in additional waiting times, thereby impacting latency. Consequently, latency is essential to take into account the scheduling order of tasks throughout the entire time span, which is referred to as the long-term QoS parameter.

\textbf{Defination 2.} (Short term QoS parameter)  Over a short period of time, the network remains stable, and some parameters that affect their QoE are solely dependent on the placement location. For instance, data accuracy and energy consumption can be computed once the placement and transmission links are determined. We refer to these parameters that affect task scheduling as short term QoS parameters.

QoE varies based on sensitivities and dimensions of QoS. From the perspective of sensitivity, different applications have different degree of susceptibility for QoS parameters.Traditional QoS typically involves hard thresholds, where applications expire once they exceed the threshold. However, in real-world scenarios, some applications may be allowed to slightly exceed the threshold. For instance, non-time-sensitive applications may tolerate a small increase in latency while still maintaining acceptable performance. Therefore, we can categorize QoS parameters into two types hard threshold and soft threshold. 

From the perspective of dimension, user satisfaction is also closely related to energy consumption, data quality etc. However, many existing studies only focus on latency as a single dimension, which may not effectively enhance user satisfaction. For analytical simplicity, we consider latency and data accuracy as different types of QoS parameters representing long term and short term QoS parameter as multi-dimensional QoS constraints. Each application has its QoS threshold denoted by ${\overset{\sim}{\mathcal{G}}}_{n}^{Qos}(t) = \left\{ d_{n}^{}(t),e_{n}^{}(t) \right\}$, where $d_n$ indicates the deadline of appication, and $e_n(t)$ represents the upper limit of data error. The preferences for QoS parameters vary for different applications. For instance, medical image processing applications have low requirements for delay with soft deadline threshold, and high requirements for accuracy with hard accuracy threshold. However, autonomous driving applications have high requirements for delay with hard deadline threshold. Once the application processing execute the deadline, the application will expire. Therefore, we use the metrics to present the application preference donated by $P_n(t)=\{W_n(t),H_n(t)\}$, where $W_{n}(t) = \left\{ w_{n}^d(t),w_{n}^e(t)\right\}$ represents the preference of QoS parameters on the scale of 0 to 1, with a higher value indicating a stronger preference and $H_{n}(t) = \left\{ h_{n}^d(t),h_{n}^e(t)\right\}$ is the indicator of the type of threshold. If it is a hard deadline and accuracy threshold, $h_{n}^d(t)=1,h_{n}^e(t)=1$.

The properties of task ${\overset{\sim}{v}}_{n,j}(t)$ are composed of task intensive type and the amount of required computing resources as ${ \{\mathbf{q}_{n,j}(t),\mathbf{r}_{n,j}(t)} \}$. $\mathbf{q}_{n,j}(t) = \{ {q_{n,j}^{cpu}(t),q_{n,j}^{gpu}(t),q_{n,j}^{i/o}(t)}\}$ is a binary vector representing the type of task, where there is a collection of three types of tasks, including CPU-intensive, GPU-intensive, and I/O-intensive. $\mathbf{r}_{n,j}(t)=\{f_{n,j}^{cpu}(t),f_{n,j}^{gpu}(t),o_{n,j}(t)\}$ is a vector of required computing resources, where $f_{n,j}^{cpu}(t)$ and ${f_{n,j}^{gpu}(t)}$ are the task FLOPs in CPU and GPU, respectively. $o_{n,j}(t)$ is I/O transmission amount. For simplicity, we only consider the major types of resources occupied, ignoring other small amounts of resources. For example, if the task is CPU-intensive, we take the CPU resource into account, while ignoring the utilization of GPU and I/O resources.
\subsection{Model of  Communication and Computation}

A binary variable $x_{j,m}^{n}(t)$ is defined to denote the scheduling decision for task ${\overset{\sim}{v}}_{n,j}(t)$. If ${\overset{\sim}{v}}_{n,j}(t)$ is assigned to node $v_m(t)$, $x_{j,m}^{n}(t)=1$; otherwise, $x_{j,m}^{n}(t)=0$. So the task ${\overset{\sim}{v}}_{n,j}(t)$ assignment strategy can be denoted by a vector $\mathbf{X}_{j}^{n}(t)= \{ x_{j,1}^{n}(t),\cdots, x_{j,m}^{n}(t),\cdots, x_{j,N}^{n}(t)\}$. And we have

\begin{equation}
\label{ex1}
\sum\limits_{m = 1}^{N}{x_{j,m}^{n}(t) = 1}.
\end{equation}
Given that the assignment decision, the data transmission time between dependent task ${\overset{\sim}{v}}_{n,j}(t)$ assigned to $v_m(t)$ and task ${\overset{\sim}{v}}_{n,j'}(t)$ assigned to $v_{m'}(t)$ can be calculated by

\begin{equation}
\label{ex2}
C_{j,j'}^{n}(t) = \left\{ \begin{matrix}
{0,~\text{if}~v_{m}(t) = v_{m'}(t)}; \\
\frac{{\overset{\sim}{e}}_{j,j^{'}}^{n}(t)}{c_{m,m^{'}}(t)},~\text{if}~v_{m}(t)~\text{and}~v_{m^{'}}(t)~\text{are in one hop};\\
\frac{{\overset{\sim}{e}}_{j,j^{'}}^{n}(t)}{c_{m,a}(t)} + \cdots + \frac{{\overset{\sim}{e}}_{j,j^{'}}^{n}(t)}{c_{b,m^{'}}(t)},\\~\text{if}~v_{m}(t)~\text{and}~v_{m^{'}}(t)~\text{are in multiple hops}.\\
\end{matrix} \right.
\end{equation}
There are three cases in $C_{j,j^\prime}^n\left(t\right)$. If the output of task ${\widetilde{v}}_{n,j}(t)$ is directly as the input to task ${\widetilde{v}}_{n,j\prime}(t)$ at the same node, the transmission delay is considered to be 0. If the output node and input node are in one hop, the transmission delay depends on the transmission rate $c_{m,m^\prime}\left(t\right)$. If $ v_m\left(t\right)$ and $v_{m^\prime}\left(t\right)$ are in multiple hops, the transmission delay will depend on data routing through intermediate nodes.

Once the task is ready to be processed in $v_m(t)$, it means all the tasks in its parent set have been completed. We define the task execution time as
\begin{equation}
\begin{split}
\label{ex3}
{T}_{j}^{n}(t) = \left\{ \begin{matrix}
{ \left| \frac{{\mathbf{q}_{n,j}(t)}^{T}\mathbf{r}_{n,j}(t)}{{\mathbf{q}_{n,j}(t)}^{T}\mathbf{F}_n(t)} \right|,~~\text{if}~{\overset{\sim}{v}}_{n,j}(t) = {\overset{\sim}{v}}_{n,0}(t)} ;\\
{\max\limits_{{\overset{\sim}{v}}_{n,\mathit{j'}}(t) \in \mathit{pre}({\overset{\sim}{v}}_{n,j}(t))}\left\{  {{T}_{j'}^{n}(t)}  + C_{j,j'}^{n}(t) \right\}} \\
+ \left| \frac{{\mathbf{q}_{n,j}(t)}^{T}\mathbf{r}_{n,j}(t)}{{\mathbf{q}_{n,j}(t)}^{T}\mathbf{F}_m(t)} \right| ,~~\text{otherwise}, \\
\end{matrix} \right.
\end{split}
\end{equation}
where $\mathit{pre}({\overset{\sim}{v}}_{n,j}(t))$ is the parent set of ${\overset{\sim}{v}}_{n,\mathit{j'}}(t) $. In order to protect the privacy of users, ${\overset{\sim}{v}}_{n,0}(t)$ can be only executed locally.

It is important to note that allocating resources to different tasks and executing them in the same node would not yield any significant benefit in our problem. Therefore, a computing node can only execute tasks of different types one at a time, while tasks of the same type are queued for execution. This approach allows for better resource utilization and efficient task execution. Thus, the finishing latency can be calculated by
\begin{equation}
\begin{split}
\label{ex4}
{T}_{j}^{n}(t) =\max\left\{{T_m(t),{\max\limits_{{\overset{\sim}{v}}_{n,\mathit{j'}}(t) \in \mathit{pre}({\overset{\sim}{v}}_{n,j}(t))}\left\{  {{T}_{j'}^{n}(t)}  + C_{j,j'}^{n}(t) \right\}}}\right\} \\
 + \left| \frac{{\mathbf{q}_{n,j}(t)}^{T}\mathbf{r}_{n,j}(t)}{{\mathbf{q}_{n,j}(t)}^{T}\mathbf{F}_m(t)} \right|,
\end{split}
\end{equation}
where $T_m(t)$ is remaining execution time of previous arranged tasks at $v_m\left(t\right)$.

The BER between task ${\overset{\sim}{v}}_{n,j}(t)$ assigned to $v_m(t)$ and task ${\overset{\sim}{v}}_{n,j'}(t)$ assigned to $v_{m'}(t)$ can be calculated by
\begin{equation}
\label{ex13}
E_{j,j'}^{n}(t) = \left\{ \begin{matrix}
{0,~\text{if}~v_{m}(t) = v_{m'}(t)}; \\
e_{m,m^{'}}(t),~\text{if}~v_{m}(t)~\text{and}~v_{m^{'}}(t)~\text{are in one hop};\\
\max\{e_{m,a}(t) , \cdots , e_{b,m^{'}}(t)\},\\~\text{if}~v_{m}(t)~\text{and}~v_{m^{'}}(t)~\text{are in multiple hops}.\\
\end{matrix} \right.
\end{equation}
There are three cases in $E_{j,j'}^{n}(t)$. If dependent tasks are arranged into the same node, the BER equal 0. If node $v_m(t)$ and node $v_{m'}(t)$ transmit data between dependent tasks within a single hop, $E_{j,j'}^{n}(t) =e_{m,m^{'}}(t)$. If the nodes are in multiple hops, the BER mainly depends on the worst channel among the hops.

Considering the dependence among tasks, the data error of task will propagate downstream. The data error rate of task ${\overset{\sim}{v}}_{n,j}(t)$ can be calculated by
\begin{equation}
\label{ex14}
{R}_{j}^{n}(t) = \left\{ \begin{matrix}
{0 ,~\text{if}~{\overset{\sim}{v}}_{n,j}(t) = {\overset{\sim}{v}}_{n,0}(t)} ;\\
{ 1-\prod\limits_{{\overset{\sim}{v}}_{n,\mathit{j'}}(t) \in \mathit{pre}({\overset{\sim}{v}}_{n,j}(t))}\left\{1-{{R}_{j'}^{n}(t)}\right\}\left\{1-E_{j,j'}^{n}(t) \right\}} ,\\\text{otherwise}, \\
\end{matrix} \right.
\end{equation}
where $1-{R}_{j'}^{n}(t)$ represents data accuracy of  the precursor of ${\overset{\sim}{v}}_{n,j'}(t)$ and $1-E_{j,j'}^{n}(t)$ is transmision accuracy.

\begin{figure}
\centering
\includegraphics[width=3.5in]{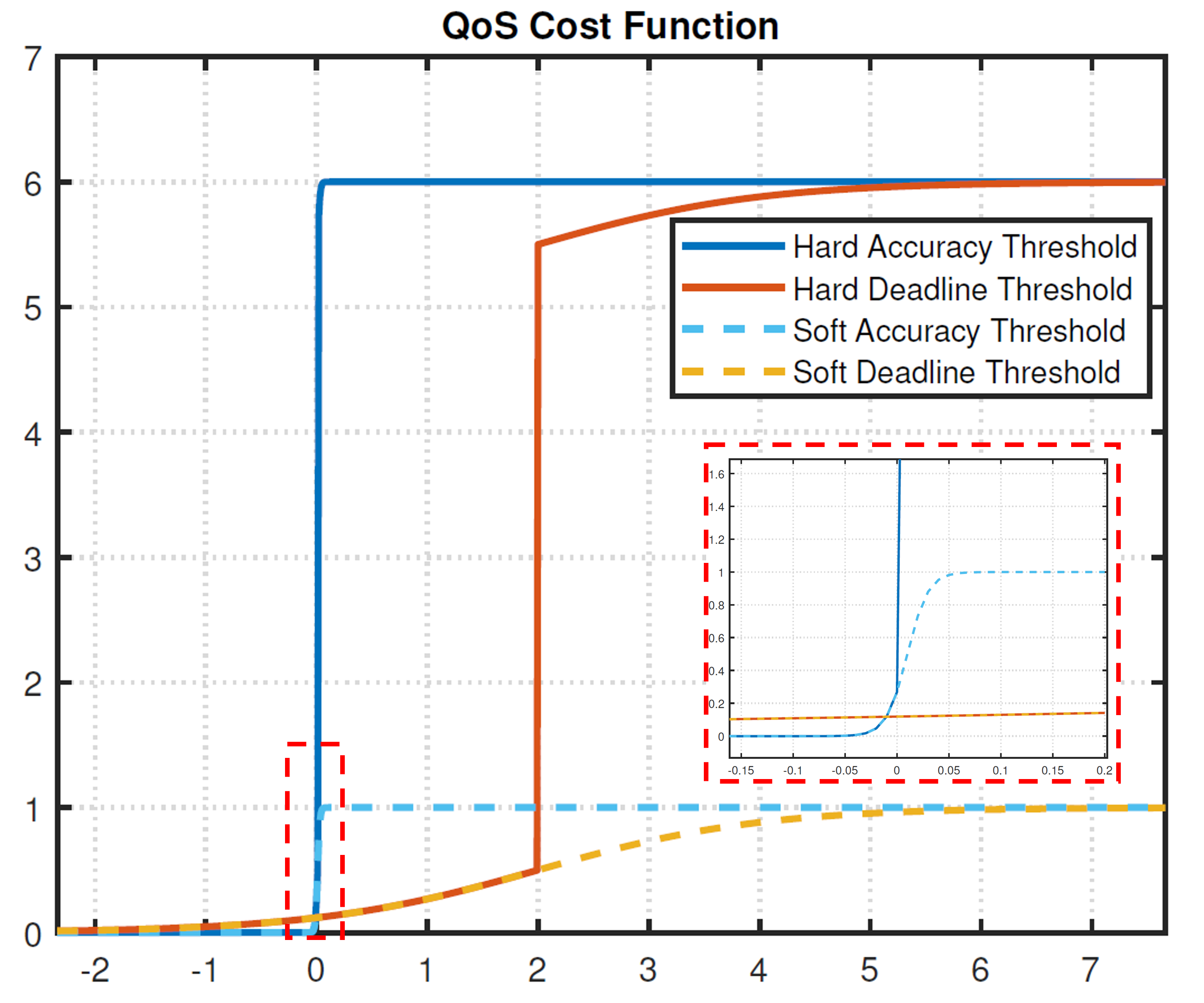}
\caption{The example of QoS cost function, where $B_d(t)$ and $B_e(t)$ are equal to 5. 
From the perspective of latency, the x-axis represents latency in seconds, and from the data accuracy perspective, the x-axis represents data error rates.}
\label{fig_2}
\end{figure}

\subsection{QoE-oriented Cost Function}
Numerous works have been proposed to predict the QoE from the input network parameters and resources, (i.e. deep learning and linear regression), so we can also get the parameters by using these methods. However, this is not our primary concern. Our focus lies in dependent task scheduling under multiple dimensions and sensitivities of QoS constraints for multiple applications over the distributed networks, aiming to maximize the overall QoE within the limited resources. In this context, it is essential to model the different dimensional QoS of dependent tasks first. The latency cost function is represented in sigmoid form as follows.
\begin{equation}
\label{ex6}
Q_{n,j}^{d}(t) =\frac{1}{1 + e^{-\frac{1}{\beta_d(t)} (T_{j}^{n}(t) - d_{n}{(t)})}}+\alpha_d(t)h_n^d(t)B_d(t),
\end{equation}
where $B_d(t)$ is the penalty factor incurred when application execution time exceeds its deadline and $\beta_d(t)$ represents the scaling factor controling the slope of the sigmoid function.

\begin{equation}
\label{ex9}
{\alpha_d(t)} =
\begin{cases}
0,&  \ T_{j}^{n}(t) - d_{n}{(t)}\leq 0,\\
{1,}&\ T_{j}^{n}(t) - d_{n}{(t)}> 0,
\end{cases}
\end{equation}
where $\alpha_d(t)$ is an indicator. If execution time exceeds deadline, $\alpha_d(t)=1$; otherwise,  $\alpha_d(t)=0$.

Data accuracy is crucial for applications that require reliable data transmission, such as medical image processing, which direct impacts on the prediction accuracy, and further influence the QoE throughout the entire process. The data accuracy degradation cost can be calculated by
\begin{equation}
\label{ex7}
Q_{n,j}^{e}(t) =\frac{1}{1 + e^{-\frac{1}{\beta_e(t)} (R_{j}^{n}(t) - e_{n}{(t)})}}+\alpha_e(t)h_n^e(t)B_e(t).
\end{equation}
Similarly,  $B_e(t)$ and $\beta_e(t)$ are the penalty factor and scaling factor, respectively.
\begin{equation}
\label{ex10}
{\alpha_e(t)} =
\begin{cases}
0,&  \ R_{j}^{n}(t) - e_{n}{(t)}\leq 0,\\
{1,}&\ R_{j}^{n}(t) - e_{n}{(t)}> 0,
\end{cases}
\end{equation}
The function curves of latency cost and data accuracy degradation cost of different sensitivities are shown in Fig. 2. The QoS cost will increase sharply when the QoS parameter is with hard threshold.  Therefore, given the preference of different QoS parameters, the QoE cost function of task ${\overset{\sim}{v}}_{n,j'}(t)$ can be modeled as follows.
\begin{equation}
\label{ex8}
QoE_{n,j}(t) = w_{n}^d{(t)}Q_{n,j}^{d}(t)+w_{n}^e{(t)}Q_{n,j}^{e}(t) .
\end{equation}

\subsection{Problem Formulation}
The objectibe is to maximize the overall QoE for all dependent tasks over the distributed networks by assigning tasks into Helping Nodes for parallel computing to tailor their QoS constraints. The optimization problem can be transformed into scheduling tasks to minimize the overall QoE cost of App Nodes. The QoE cost is determined upon the completion of applications. Therefore, the optimization problem can be formulated as follows.
\begin{equation}
\label{ex9}
\min\limits_{\mathbf{X}(t)}\frac{\sum\limits_{n = 1}^{N}{v_{n}^{'}(t) QoE_{n,J_{n}}}(t)}{\sum\limits_{n = 1}^{N}{v_{n}^{'}(t)}},
\end{equation}
s. t. \begin{center}
(1),\ (3),\ (4),\ (6).
\end{center}

$\mathbf{X}(t)$ is the matrix of assignment strategy for all tasks. Constraint (1) ensures that each task can only be assigned to a node. Constraints (3-6) guarantee that both computing and communication depend on the logical sequence of DAG applications.

\textbf{Theorem 1.} Problem (12) is Np-hard.

\textbf{Proof.} To prove this problem is Np-hard, we reduce it into Shop Scheduling Problem, which is a well-known Np-hard problem \cite{ref10}.
Shop Scheduling is a classic combinatorial optimization problem, typically described as a set of jobs that need to be processed on a set of machines. Each job comprises a sequence of subtasks, each with different processing times. The objective is to find a scheduling scheme that minimizes the total completion time for all jobs.
We consider a special case of our problem, where each node can only process a task of a application.  If the special case is Np-hard, the original problem is also Np-hard. 

we construct a corresponding instance for the Shop Scheduling Problem as follows:
\begin{itemize}
\item[1)] Machines: Create a set of machines, where each machine corresponds to a computing node available for task execution in our problem.

\item[2)] Jobs: Each DAG application is treated as a job in the Shop Scheduling instance.

\item[3)] Tasks: Map the tasks of each application to subtasks within the corresponding job in Shop Scheduling. The processing times of these subtasks represent the execution and transimition time required for task execution on the associated computing node. Importantly, each subtask has its associated machine, indicating where it should run.

\item[4)] Dependencies: Reflect the task dependencies from our problem in the Shop Scheduling instance. If one task depends on the completion of another, this is translated into task dependencies between the corresponding subtasks in Shop Scheduling Problem.

\item[5)] Equivalence:  The objective of Shop Scheduling Problem is to minimize the makespan, which is the total time required to complete all jobs. This aligns with our problem objective of minimizing the average QoE cost for all tasks across multiple DAGs.
\end{itemize}

Since this variant of the Shop Scheduling Problem is still NP-hard, our reduction establishes that the adapted original problem is also NP-hard. This completes the proof.
\section{Hierarchical Multi-queue Task Scheduling Algorithm}
\begin{figure}
\centering
\includegraphics[width=3.5in]{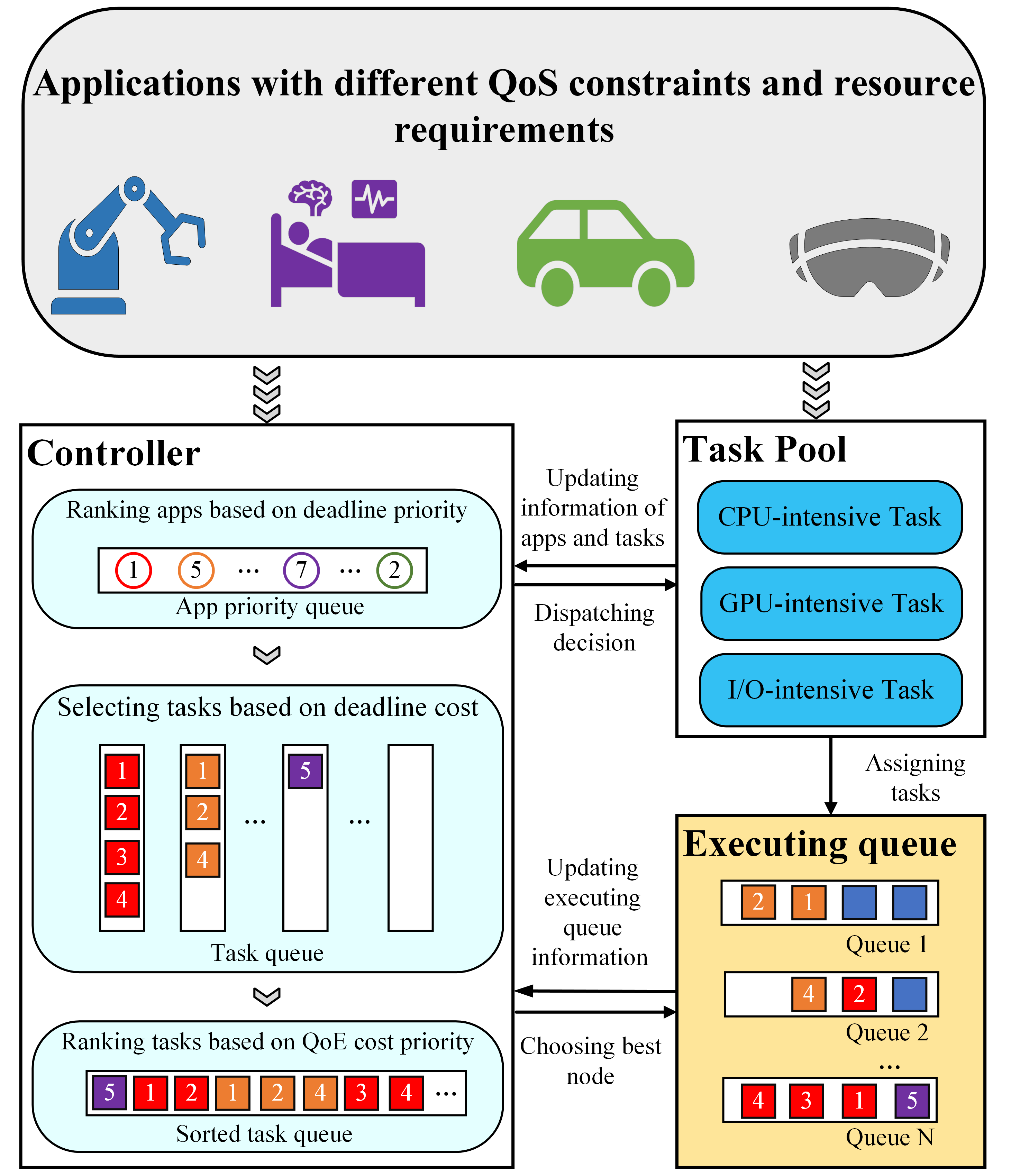}
\caption{Illustritive diagram of HMTSA.}
\label{fig_3}
\end{figure}
Scheduling multiple dependent tasks poses a challenge of high computational complexity, especially with the exponential growth of the solution space as the number of applications increases. Moreover, the design of scheduling strategy is more complicated due to the unknown number and arrival time of dependent tasks. In order to efficiently schedule dependent tasks, task prioritization is the most important issue. However, DAG applications scheduling in the distributed networks make the challenge extend beyond considering only the logical order, communication, and computational workloads, as is common in scheduling single DAG application within specified deadlines. Here, we also grapple with the added complexities arising from the priorities of other DAG applications and multiple dimensional QoS constraints.

To tackle the aforementioned challenges, we propose a heuristic Hierarchical Multi-queue Task Scheduling Algorithm (HMTSA) as shown in Fig. 3. Initially, the dependent tasks with different QoS constraints and resource requirements are categorize according their resource-intensive requirement. Subsequently, an initial application and task prioritization strategy based on bottom level is proposed as outlined in Algorithm 1 \cite{ref27}. After ranking all tasks and applications, these tasks are organized into relative task pools, awaiting for assignment in order. Note that different task pool will perform its own task scheduling strategy in parallel, because applications for different resource requirements can not affect each other's performance for individualized task scheduling strategies within distinct task pools. Finally, a hierarchical multi-queue task scheduling algorithm is proposed based on the minimum remaining time as outlined in Algorithm 2. In this section, we provide a detailed explanation of the algorithm, concluding with an analysis of its computational complexity.
\subsection{Task Prioritization}
In order to solve multiple DAG application scheduling problem with multi-dimensional QoS constraints, priorities in latency and data accuracy dimensions should be take into account in our problem. From the latency perspective, DAG applications inherently possess natural logical orders. Relying solely on a traditional task scheduling strategy based on deadlines is not effective due to distinct transmission and computing workloads of various applications and tasks. Therefore, the total computation time and communication time of each task ${\overset{\sim}{v}}_{n,j}(t)$ can be estimated based on the longest path to the end node as follows. 

\begin{equation}
\begin{split}
\label{ex5}
T_{j}^{'n}(t) = \left\{ \begin{matrix}
{0,~~\text{if}~{\overset{\sim}{v}}_{n,j}(t) = {\overset{\sim}{v}}_{n,J_{n}}(t)}; \\
{\max\limits_{{\overset{\sim}{v}}_{n,\mathit{j'}}(t) \in \mathit{suc}({\overset{\sim}{v}}_{n,j}(t))}\left\{  {T_{j'}^{'n}(t)} + \frac{{\overset{\sim}{e}}_{j,j'}^{n}(t)}{\overset{-}{c}(t)} \right\}} \\
+ \left| \frac{{\mathbf{q}_{n,j'}(t)}^{T}\mathbf{r}_{n,j'}(t)}{{\mathbf{q}_{n,j'}(t)}^{T}{\overset{-}{\mathbf{F}}(t)}} \right| ,~~\text{otherwise}, \\
\end{matrix} \right.
\end{split}
\end{equation}
where $\bar{c}$ and $\bar{\mathbf{F}}(t)$ indicate the average transmission rate and computing capability among all nodes, respectively. $suc({\widetilde{v}}_{n,j}(t))$ is the children set of ${\widetilde{v}}_{n,j}(t)$. 
Moreover, the task priority in the latency dimension can be calculated by
\begin{equation}
\begin{split}
\label{ex14}
Q_{n,j}^{'d}(t) =\frac{w_{n}^d{(t)}}{1 + e^{-\frac{1}{\beta_d(t)} (T_{j}^{'n}(t) - d'_{n,j}{(t)})}} +\alpha_d(t)w_{n}^d{(t)}h_n^d(t)B_d(t),
\end{split}
\end{equation}
where $d'_{n,j}(t){= d}_{n}^{}(t) - {T}^{n}(t)$ represents the remaining time of the application ${\overset{\sim}{\mathcal{G}}}_{n}(t)$ and $T_n(t)$ is the executing time during application processing. In the initial stage of application and task priority ranking, $T^n(t)=0$, we can get the initial task priority in the latency dimension as outlined in Algorithm 1. As scheduling progresses,  $T^n(t)$ is updated to the maximum execution time of the assigned tasks of application ${\overset{\sim}{\mathcal{G}}}_{n}(t)$.
\begin{algorithm}[H]
\caption{Initial Application and Task Priority Ranking}\label{alg:alg1}
\begin{algorithmic}
\STATE 
\STATE \textbf{Input } The set of generating DAG applications, average transmission rate, computing capability and BER.
\STATE \textbf{Output } The pool of all task  $\mathbb{T}$ and the queue of application $\mathbb{A}$ in a descending order .
\STATE \textbf{Initialization:} task pooling $\mathbb{T}=\emptyset$, application $\mathbb{A}=\emptyset$, $T_{j}^{'n}(t)  =-\infty$,  $R_{j}^{'n}(t)  =-\infty$.
\STATE \textbf{for} $n =1 \to N$ \textbf{do}
\STATE \hspace{0.2cm} \textbf{if} $v'_n(t)=1$ \textbf{do}
\STATE \hspace{0.4cm} $T_{J_n}^{'n}(t) =0$, $R_{J_n}^{'n}(t) =0$;
\STATE \hspace{0.4cm} \textbf{for} each $\overset{\sim}{v}_{n,j}(t) \in \overset{\sim}{v}(t)$ in topological sorting order \textbf{do}
\STATE \hspace{0.6cm} Calculate all $T_{j}^{'n}(t), Q_{n,j}^{'d}(t), R_{j}^{'n}(t), Q_{n,j}^{'e}(t)$ according to Eq. (13-16);
\STATE \hspace{0.6cm} $\mathbb{T}_{n,j}\gets\{Q_{n,j}^{'d}(t)+Q_{n,j}^{'e}(t)\}$;
\STATE \hspace{0.6cm} $\mathbb{T}_{n}\gets \mathbb{T}_{n}+\mathbb{T}_{n,j}$;
\STATE \hspace{0.4cm} \textbf{end for}
\STATE \hspace{0.4cm} $\mathbb{T}\gets \mathbb{T}+\mathbb{T}_{n}$;
\STATE \hspace{0.4cm} $\mathbb{A}_n\gets \{n,Q_{n,0}^{'d}(t)\}$;
\STATE \hspace{0.4cm} $\mathbb{A}\gets \mathbb{A}+\mathbb{A}_n$;
\STATE \hspace{0.2cm} \textbf{end if}
\STATE \textbf{end for}
\STATE $\mathbb{T}.sort(Q_{n,j}^{'d}(t)+Q_{n,j}^{'e}(t))$;
\STATE $\mathbb{A}.sort(Q_{n,0}^{'d}(t))$.
\end{algorithmic}
\label{alg1}
\end{algorithm}
Similarly, from the data accuracy perspective, the data error rate of task ${\overset{\sim}{v}}_{n,j}(t)$ can be estimated by
\begin{equation}
\label{ex15}
{R}_{j}^{'n}(t) = \left\{ \begin{matrix}
{0 ,~\text{if}~{\overset{\sim}{v}}_{n,j}(t) = {\overset{\sim}{v}}_{n,J_{n}}(t)};\\
{ 1-\prod\limits_{{\overset{\sim}{v}}_{n,\mathit{j'}}(t) \in \mathit{suc}({\overset{\sim}{v}}_{n,j}(t))}\frac{1-{{R}_{j'}^{'n}(t)}}{1-\bar{e}(t) }} ,\text{otherwise}, \\
\end{matrix} \right.
\end{equation}
where $\bar{e}(t)$ represents the average BER of the communication channel. 
And the task priority in the data accuracy dimension can be calculated by
\begin{equation}
\begin{split}
\label{ex16}
Q_{n,j}^{'e}(t) =\frac{w_{n}^e{(t)}}{1 + e^{-\frac{1}{\beta_e(t)}\{ 1-(1-R_{j}^{n}(t))(1-R_{j}^{'n}(t)) - e_{n}{(t)}\}}}\\ 
+ \alpha_e(t)w_{n}^e{(t)}h_n^e(t)B_e(t),
\end{split}
\end{equation}
where $1-(1-R^{n}(t))(1-R_{j}^{'n}(t))$ is the estimated data error rate. In the initial stage of prioritizing tasks, $R^n(t)=0$. As scheduling progresses, $R^n(t)$ is updated to the maximum data error rate of the assigned tasks of app ${\overset{\sim}{\mathcal{G}}}_{n}(t)$. 

A larger value of $Q_{n,j}^{'d}(t)$ and $Q_{n,j}^{'e}(t)$ suggests that more time or better communication channel is required for task processing, indicating a higher priority of the task. Note that, the value of the first task of each application is the priority of the application. 
According to the equation (13)-(16), Algorithm 1 obtains the initial priority of each task using the dynamic programming algorithm, and puts it into a sequence according to the descending order, and obtains the priorities of all tasks and applications.
\begin{algorithm}[H]
\caption{Multi-queue Task Scheduling}\label{alg:alg2}
\begin{algorithmic}
\STATE 
\STATE \textbf{Input} The pooling of all task  $\mathbb{T}$, the queue of application $\mathbb{A}$.
\STATE \textbf{Output} The task scheduling results $\Phi$. 
\STATE \textbf{Initialization:} application queue $q_a=[]$, task priority queues $q_{a,i}=[]$, sorted task queue $q_t = []$, executing queues of each node $q_n= []$.
\STATE \textbf{while} $\mathbb{T} \neq \emptyset$ \textbf{do} 
\STATE \hspace{0.2cm} \textbf{for} $i=1\gets u$ \textbf{do} // $u$ is the number of scheduling application 
\STATE \hspace{0.4cm} $q_a\gets q_a +\mathbb{A}[i][1]$ // getting the label of the application whose latency priority ranks $i$th
\STATE \hspace{0.4cm} Calculate $M_i$ according to Eq. (17);
\STATE \hspace{0.4cm} \textbf{for} $x=1\gets$ $M_i$ \textbf{do}
\STATE \hspace{0.6cm} $q_{a,i}.append(\mathbb{T}_{q_a[i]}[x])$ // getting the top $x$th tasks of $\mathbb{T}_{q_a[i]}$ in latency priority;
\STATE \hspace{0.6cm} $\mathbb{T} = \mathbb{T}-\mathbb{T}_{q_a[i]}[x]$;
\STATE \hspace{0.4cm} \textbf{end for}
\STATE \hspace{0.4cm} $q_t \gets q_t + q_{a,i}$;
\STATE \hspace{0.4cm} \textbf{if} $\mathbb{T}_{q_a[i]} =\emptyset$ \textbf{do}
\STATE \hspace{0.6cm} $\mathbb{A} = \mathbb{A}-\mathbb{A}[i]$;
\STATE \hspace{0.4cm} \textbf{end if}
\STATE \hspace{0.4cm} $\mathbb{A}[i]=max(\mathbb{T}_{q_a[i]})$; // updating $\mathbb{A}$
\STATE \hspace{0.4cm} $\mathbb{A}=\mathbb{A}.sort(\mathbb{A}[i][2])$; // ranking the application according to the updated latency priority in application level
\STATE \hspace{0.2cm} \textbf{end for}
\STATE \hspace{0.2cm}  $q_t.sort()$; // adjusting the order based on QoE priority in task level
\STATE \hspace{0.2cm} \textbf{for} $q_{t,s}\in q_t$ \textbf{do}
\STATE \hspace{0.4cm} Choose the node $n$ getting minimal QoE cost according to Eq. (2-6), (11);
\STATE \hspace{0.4cm} $q_n.append(q_{t,s})$;
\STATE \hspace{0.4cm} $q_t.remove(q_{t,s})$;
\STATE \hspace{0.4cm} Update $\Phi$;
\STATE \hspace{0.2cm} \textbf{end for}
\STATE \textbf{end while}
\end{algorithmic}
\label{alg2}
\end{algorithm}
\subsection{Hierarchical Multi-queue Task Scheduling}
Since the short term parameter depend on the location of placement and will not change over time, while the long term factors mentioned in Definition 1 will change due to the order of placement. Therefore, we propose a hierarchical multi-queue task scheduling algorithm as illustrated in Fig. 3. The queues are divided into scheduling queues and executing queues. The hierarchical scheduling queues consist of the application queue, task priority queue and the sorted task queue. Each node has its own executing queue, where tasks positioned further back in the executing queue need to wait for the completion of tasks positioned ahead in the queue in a non-preemptive scenario. Our algorithm necessitates the maintenance of these queues and the pool to record the remaining tasks and the remaining application priorities. The specific scheduling process is outlined in Algorithm 2.
\begin{figure}[t]
\centering
\includegraphics[width=3.6in]{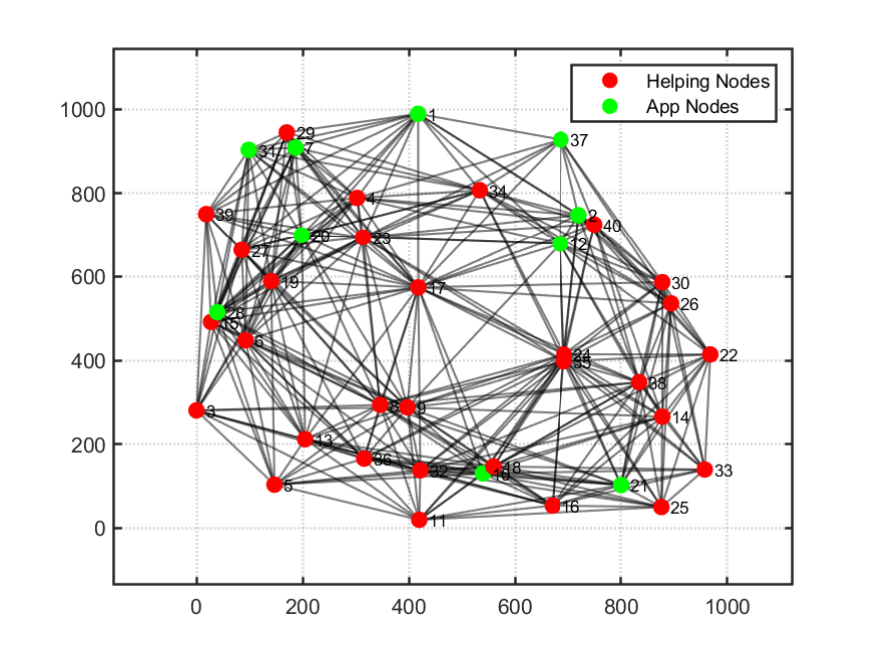}
\caption{Simulation scenario. }
\label{fig_3}
\end{figure}

Algorithm 2 is the core of our proposed task scheduling algorithm. The input of Algorithm 2 is the task pool with all prioritilizing tasks and application outputing of Algorithm 1. Firstly, in order to ensure to assign more tasks of an application with higher latency priority to reduce latency cost in the application level, it performs application ranking according to the application priority in latency dimension. Assume that the priority of application ${\overset{\sim}{\mathcal{G}}}_{n}(t)$ is $Q_n^{'d}(t)$ in the latency priority, and the ratio of  application priority queue is $k$, and the ratio of task queue is $o$. As shown in Fig. 3, the ratio of app priority queue refers to the proportion of applications entering the queue  compared to the total number of applications and the ratio of task queue refers to the proportion of tasks of each application entering the task queue during each round of scheduling event, respectively. To determine the task assignment for each application, we normalize their priorities. The number of selecting application (the length of application priority queue) is $u=\lfloor k\sum\limits_{n = 1}^{N}{v_{n}^{'}(t)}\rfloor$.  The number of tasks assigned to each application (the length of task queue of each node) according to the equation as follows.
\begin{equation}
\label{ex12}
{M}_n=\left\lfloor\frac{oJ_nQ_{n}^{'d}(t)}{\sum_{u^\prime=1}^{u}Q_{u\prime}^{'d}(t)}\right\rfloor.
\end{equation}

After determining the number of tasks for each application, tasks are selected based on their priority in the latency dimension and placed into the respective task queues. Subsequently, at the task level, these tasks are prioritized according to the task priority of QoE cost and enter the sorted task queue. This strategy alters the assignment order, giving higher priority to tasks with strict BER requirements. Therefore, high priority tasks in latency dimension are assigned to slightly less optimal nodes, enabling them to achieve satisfactory performance while preserving the excellent performance of the best node for tasks with stringent BER requirements. Finally, tasks in the sorted task queue based on QoE cost priority are assigned to choose the best node with the minimum QoE cost. After a round of scheduling, the priority of all applications and tasks is updated, and the assigned tasks and applications are deleted. This process is repeated until all tasks have been assigned.
\subsection{Complexity Analysis}
The computational complexity of our HMTSA is $O(N(J+M)+JN^3)$, which is determined by considering the most complex operation. We assume that there are $N$ nodes in the network, each requiring application processing consisting of $J$ tasks and having $M$ edges. The computational complexity is comprised of two main components: task prioritizing and multi-queue task scheduling.
\begin{figure*}[t]
    \centering
     \setstretch{0.0}
\vspace{0cm}
\renewcommand{\sfdefault}{ptm}
    \subfloat[]{\includegraphics[width=0.32\linewidth]{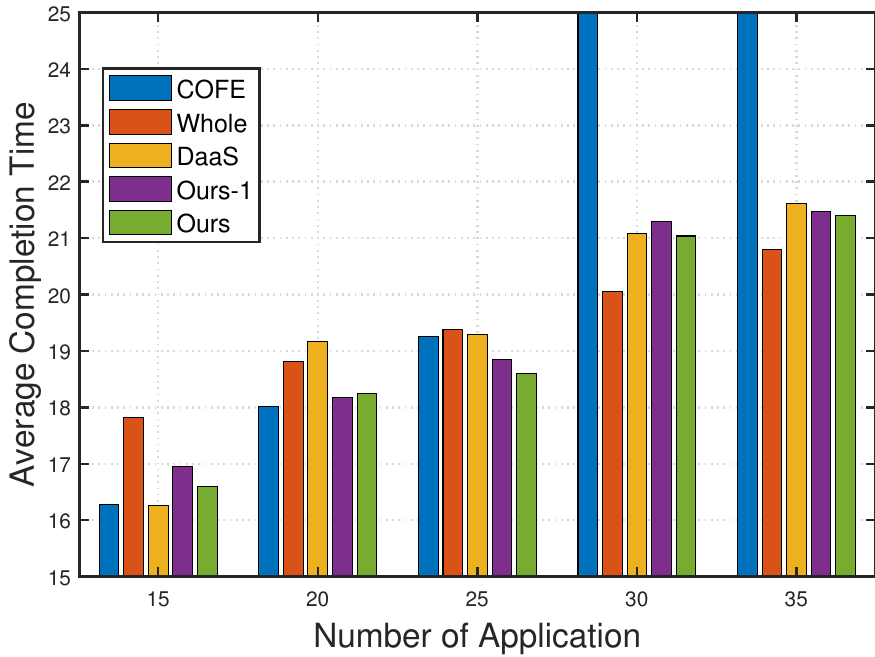}\captionsetup{font={sf,small}}}
   \hspace{0.01\linewidth} 
    \subfloat[]{\includegraphics[width=0.32\linewidth]{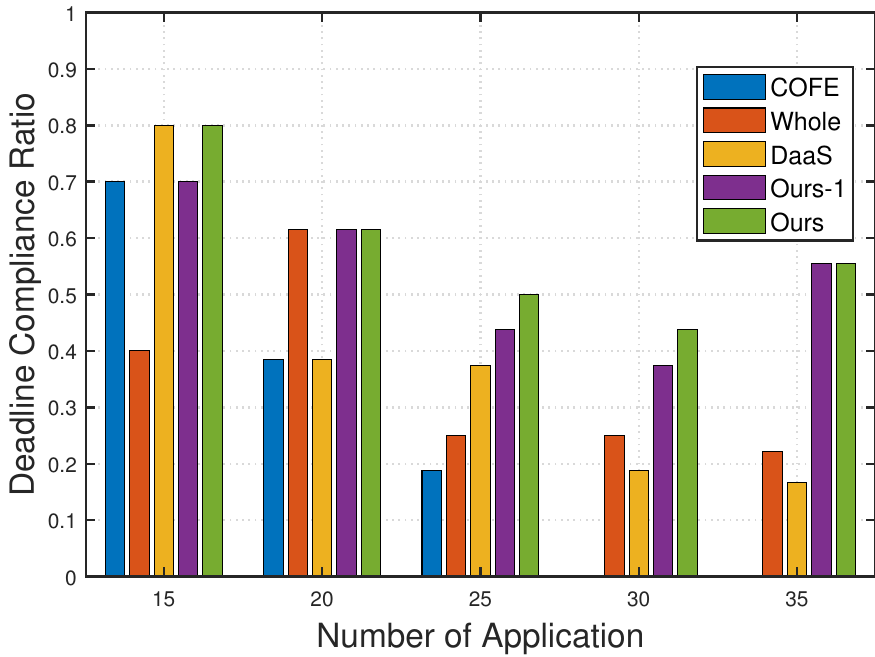}\captionsetup{font={sf,small}}}
\hspace{0.01\linewidth} 
    \subfloat[]{\includegraphics[width=0.32\linewidth]{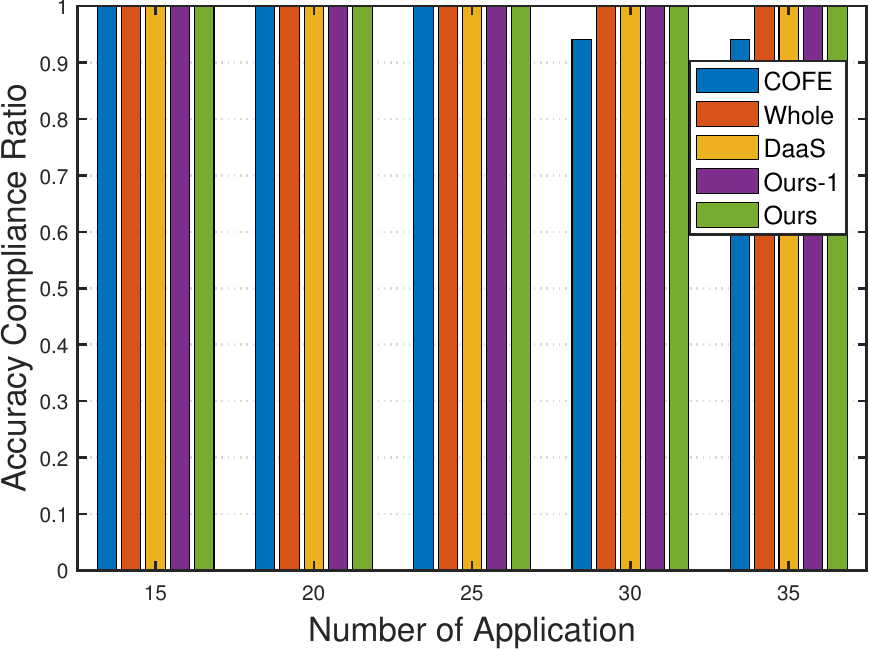}\captionsetup{font={sf,small}}}
     \vspace{0.1cm} 
    \subfloat[]{\includegraphics[width=0.32\linewidth]{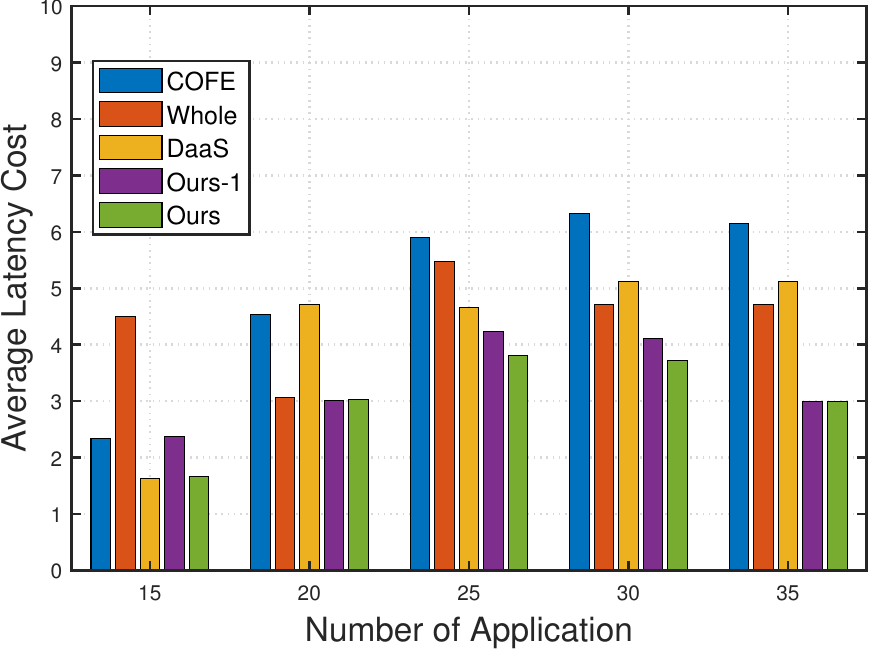}\captionsetup{font={sf,small}}}
\hspace{0.01\linewidth} 
    \subfloat[]{\includegraphics[width=0.32\linewidth]{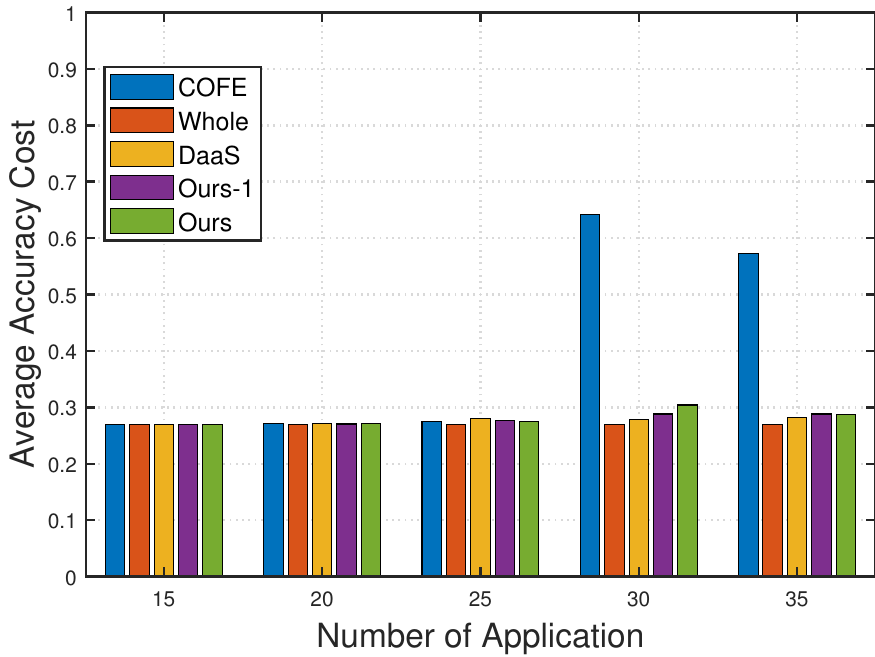}\captionsetup{font={sf,small}}}
\hspace{0.01\linewidth} 
    \subfloat[]{\includegraphics[width=0.32\linewidth]{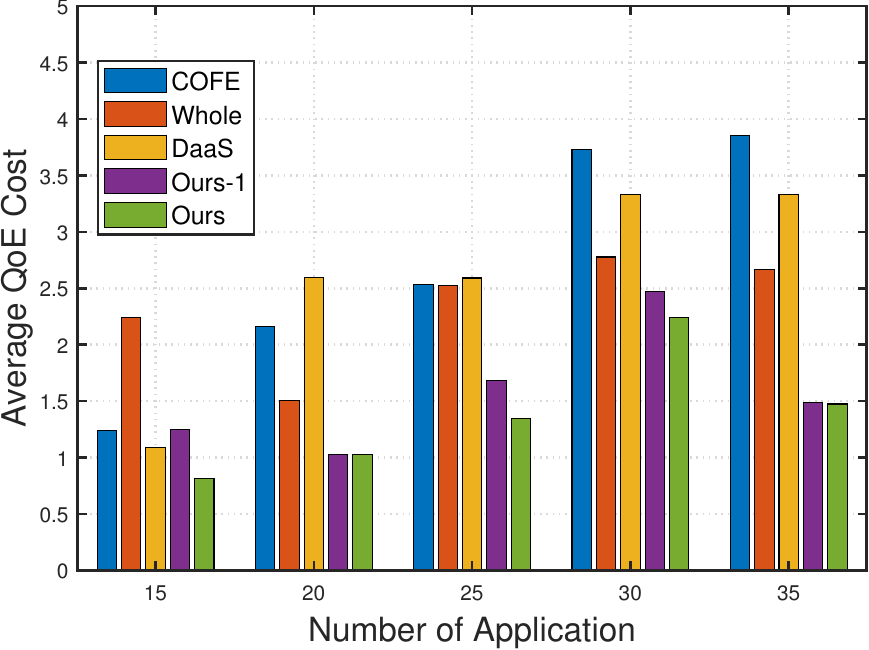}\captionsetup{font={sf,small}}}
    
    \caption{Impact of number of application. (a) Average completion time versus the number of application. (b) Deadline compliance ratio versus the number of application. (c) Accuracy compliance ratio versus the number of application. (d) Average latency cost versus the number of application. (e) Average accuracy cost versus the number of application. (f) Average QoE cost versus the number of application.}
\end{figure*}

The first part involves using the dynamic programming algorithm to obtain the task and application in Algorithm 1, and its complexity is therefore recognized as $O(N(J+M))$. The second part relies on three nested loops to determine the best node for each task. In the most complex scenario, we assume that only one task is scheduled in each round. Hence, the complexity of the first loop is $O(JN)$ to decide how many tasks should be assigned. The complexity of the second loop is $O(N)$, which identifies the most urgent application. The third loop is used to find the best node to minimize the QoE cost of the task, and its complexity can be recognized as $O(N)$.The complexity of task scheduling is $O(JN * N * N) = O(JN^3)$. Consequently, the computational complexity of our algorithm is $O(N(J+M)+JN^3)$.
\section{Performance Evaluation}
In this section, we conduct extensive experiments to evaluate the performance of HMTSA using multiple metrics. The simulation setup and results are described in detail in the following subsection.

\begin{figure*}[t]
\vspace{0cm} 
    \centering
     \setstretch{0.0}
\renewcommand{\sfdefault}{ptm}
    \subfloat[]{\includegraphics[width=0.32\linewidth]{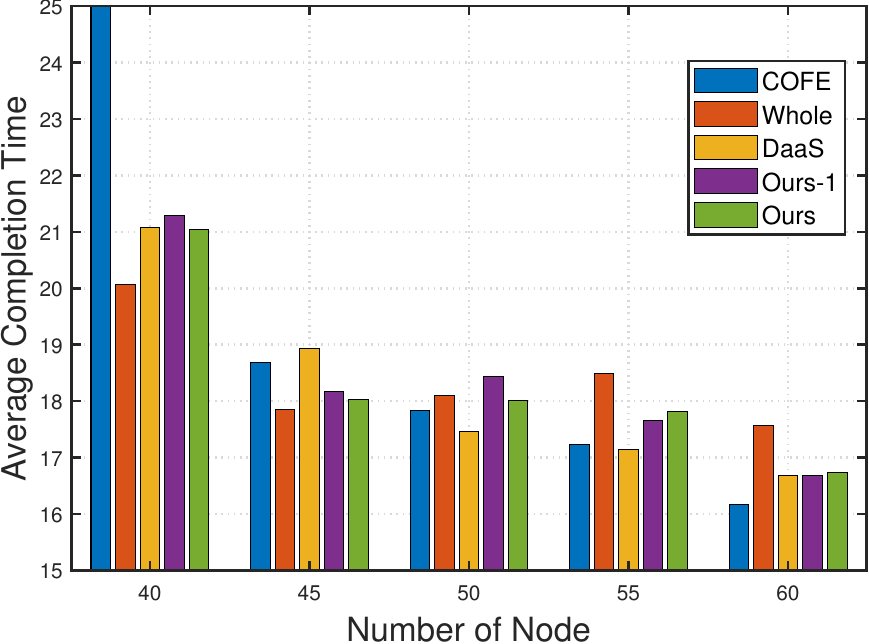}\captionsetup{font={sf,small}}}
   \hspace{0.01\linewidth} 
    \subfloat[]{\includegraphics[width=0.32\linewidth]{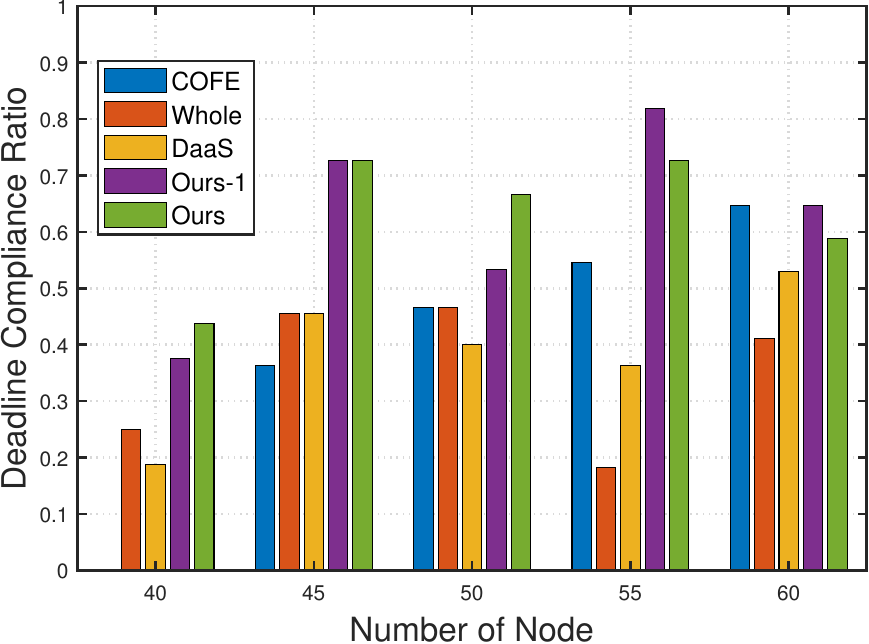}\captionsetup{font={sf,small}}}
\hspace{0.01\linewidth} 
    \subfloat[]{\includegraphics[width=0.32\linewidth]{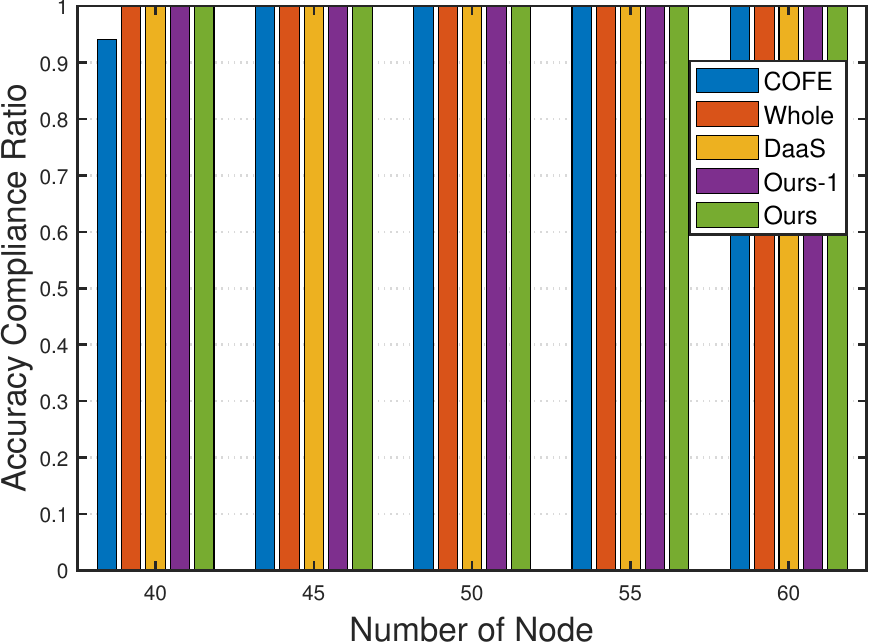}\captionsetup{font={sf,small}}}
     \vspace{0.1cm} 
    \subfloat[]{\includegraphics[width=0.32\linewidth]{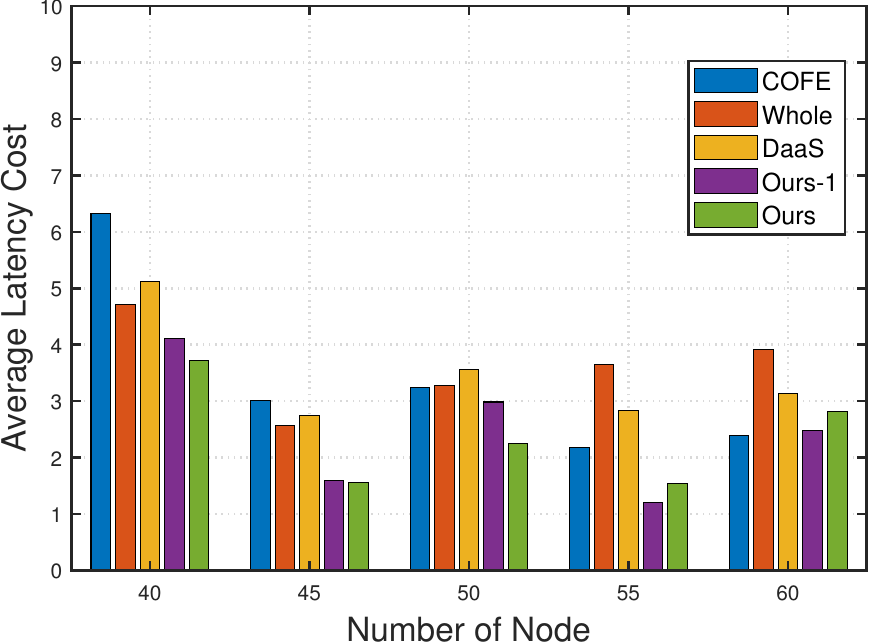}\captionsetup{font={sf,small}}}
\hspace{0.01\linewidth} 
    \subfloat[]{\includegraphics[width=0.32\linewidth]{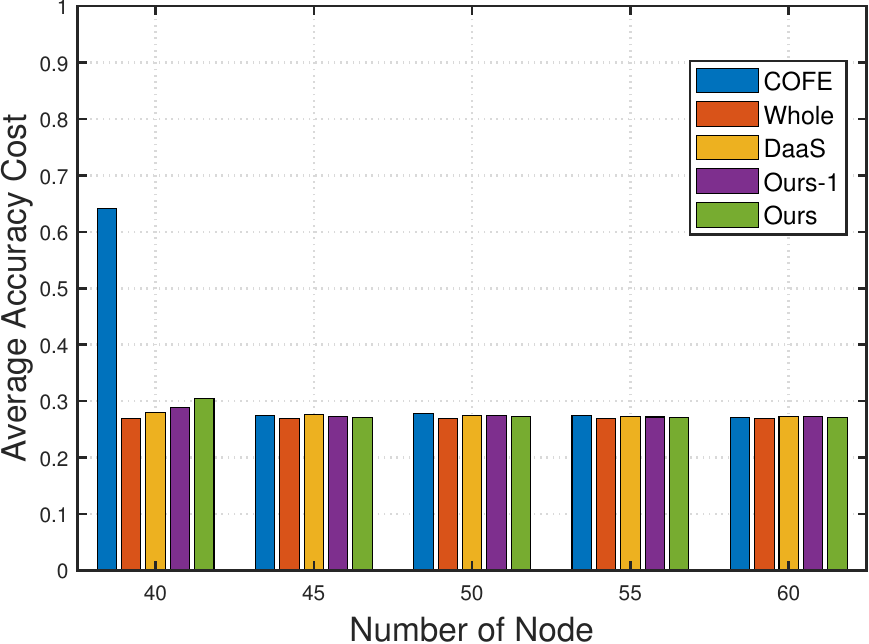}\captionsetup{font={sf,small}}}
\hspace{0.01\linewidth} 
    \subfloat[]{\includegraphics[width=0.32\linewidth]{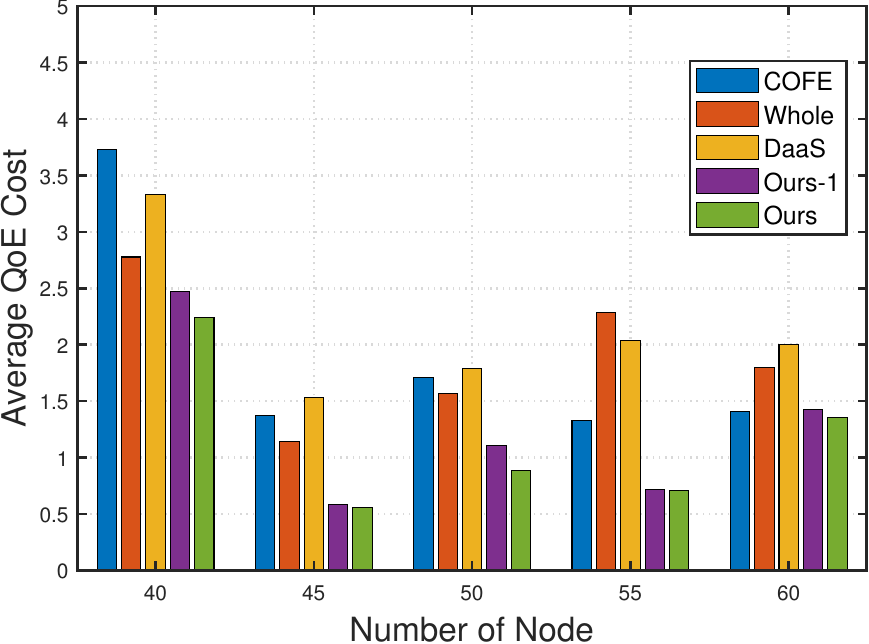}\captionsetup{font={sf,small}}}
    
    \caption{Impact of number of node. (a) Average completion time versus the number of node. (b) Deadline compliance ratio versus the number of node. (c) Accuracy compliance ratio versus the number of node. (d) Average latency cost versus the number of node. (e) Average accuracy cost versus the number of node. (f) Average QoE cost versus the number of node.}
\end{figure*}
\subsection{Simulation Setup}
We consider a square area of 1000*1000 $m^2$, where 40 nodes are randomly distributed with computing capacity by default as shown in Fig. 4. The computing capabilities of these nodes fall within the range of 5 to 10 GHz. Each node can communicate with its neighboring nodes within a distance of 500 meters, utilizing variable transmission rates between 0 and 20 MB/s. Additionally, the Bit Error Rate (BER) of transmission link is randomly selected from the set \{$10^{-4}$, $10^{-5}$, $10^{-6}$, $10^{-7}$\}. Notably, a subset of these nodes (App Nodes) will generate applications with dependent tasks, indicated by the green labels.

We randomly generate applications with dependent tasks, and set a structure by following characteristics: 
The task number of application is randomly generated in the set of \{16, 17, 18, 19, 20\}. 
The branch is selected from the set \{2, 3, 4, 5\}, meaning the application will randomly generate branches below 5.
The computing workload is set to [5, 10] GHz, and the transmission workload is randomly selected in [0.1, 0.5] MB. The deadline of application is randomly generated from [15, 20] s and the accuracy threshold of application is randomly selected from the set \{$10^{-2}$, $10^{-3}$, $10^{-4}$\}. The ratio of application with hard threshold is set as 0.5 by default and $B_d(t)$ and $B_e(t)$ are set to 10. 

In order to verify the superiority of our algorithm,we set the ratio of application priority queue $k$ and the ratio of task queue $o$ as a constant, 1/4 and 1 respectively. And four baselines are set in our simulation environment, which are described as follows.
\begin{itemize}
\item[1)] COFE: This baseline is modified based on COFE to tailor it to our specific scenario\cite{ref27}. Task scheduling is triggered only upon the completion of a task. When multiple tasks require scheduling simultaneously, they are initially prioritized based on their deadlines and subsequently sorted by their task priorities. Additionally, the node selection principle has shifted from selecting the node with the shortest completion time to minimizing the QoE, aligning with our problem.

\item[2)] Daas: This baseline is modified based on DaaS to be tailored to our specific scenario \cite{r4}. It involves estimation task QoE priority of all application and assignment of tasks to nodes according to the priority without updating the actual remaining time.

\item[3)] Whole : All applications are sorted based on their estimated latency degradation costs. Subsequently, applications are sequentially assigned to the nodes with the shortest computation duration in the sorted order.

\item[4)] Ours1: This algorithm is designed to compare our algorithm without considering hierarchical sorting. It ranks applications based on QoE cost.  
\end{itemize}

In our simulations, the performance of algorithms is evaluated using six metrics, which are defined as follows.
\textit{Average completion time}: The average makespan of all applications.
\textit{Deadline compliance ratio}: The completion rate of tasks with hard deadline constraints.
\textit{Accuracy compliance ratio}: The completion rate of tasks with hard data accuracy constraints.
\textit{Average latency cost}: The average makespan cost of all applications.
\textit{Average accuracy cost}: The average data accuracy cost of all applications.
\textit{Average QoE cost}: The average QoE cost of all applications.
\subsection{Simulation Results}

\subsubsection{Impact of Number of Application}
In order to test the performance of our proposed algorithm, we adjust the number of application in the mesh network and observe the performance using the six metrics as depicted in Fig. 5. It can be observed that the average completion time, average latency cost, and average QoE cost increase with the rising number of applications as shown in Fig. 5(a)(d)(f). Conversely, the deadline compliance ratio decreases as the number of application increases, as illustrated in Fig. 5(b). Additionally, Fig. 5(c) and 5(e) reveal that the accuracy compliance ratio and average accuracy cost remain relatively stable, with the exception of the values when the number of applications reaches 30 and 35 for the COFE algorithm. The reason is that these methods tend to choose the location of the precursor task node when offloading tasks, which does not increase the accuracy cost without transmission if the latency priority does not exceed the hard deadline threshold.

We observe that our algorithm consistently outperforms the other methods, especially when the number of applications is big as shown in Fig. 5. COFE and DaaS demonstrate good performance when the number of application is 15. However, both of the methods experience significant performance degradation as the number of applications increases. In contrast, the Whole method exhibits improved performance. This differences in performance can be attributed to the following factors: COFE focuses solely on the current situation when deciding which tasks to offload. Conversely, DaaS assigns all tasks a priority at the beginning without making rectification during task scheduling. In contrast, our algorithm updates task priorities dynamically during scheduling. For instance, if an application has been assigned numerous tasks and its priority may drop below that of applications without assigned tasks, our algorithm can rectify this by updating priorities accordingly.
\begin{figure*}
    \centering
     \setstretch{0.0}

\renewcommand{\sfdefault}{ptm}
    \subfloat[]{\includegraphics[width=0.32\linewidth]{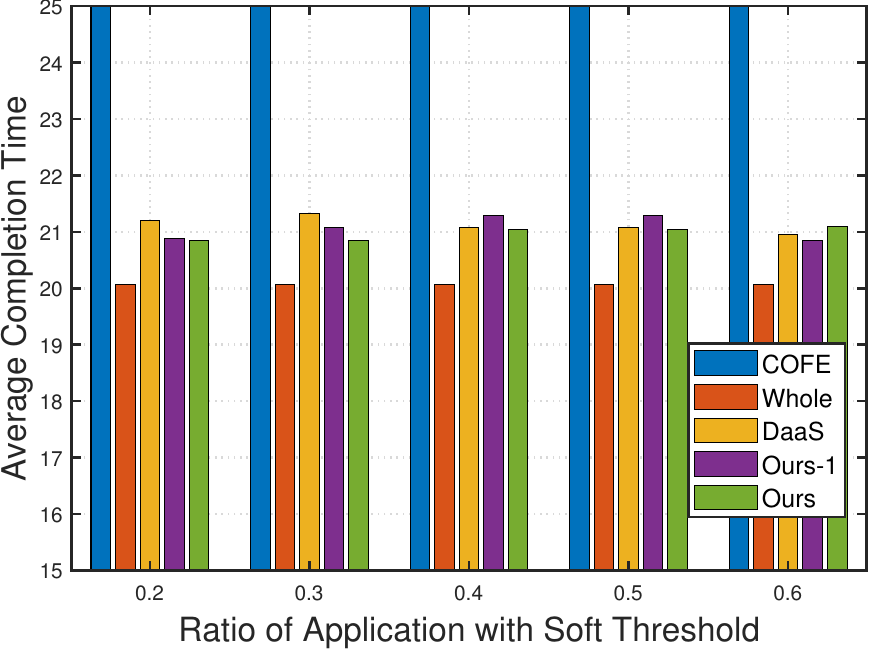}\captionsetup{font={sf,small}}}
   \hspace{0.01\linewidth} 
    \subfloat[]{\includegraphics[width=0.32\linewidth]{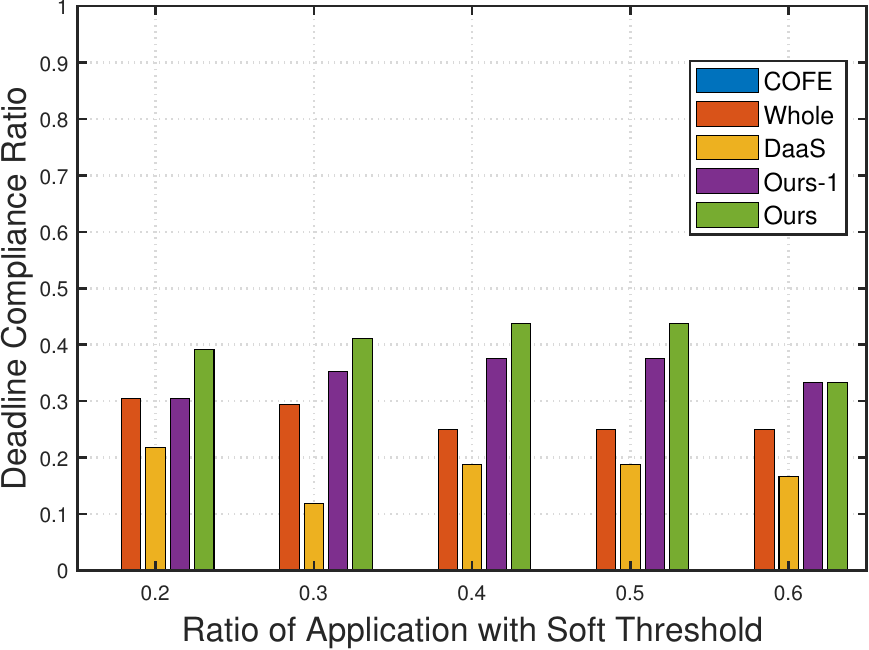}\captionsetup{font={sf,small}}}
\hspace{0.01\linewidth} 
    \subfloat[]{\includegraphics[width=0.32\linewidth]{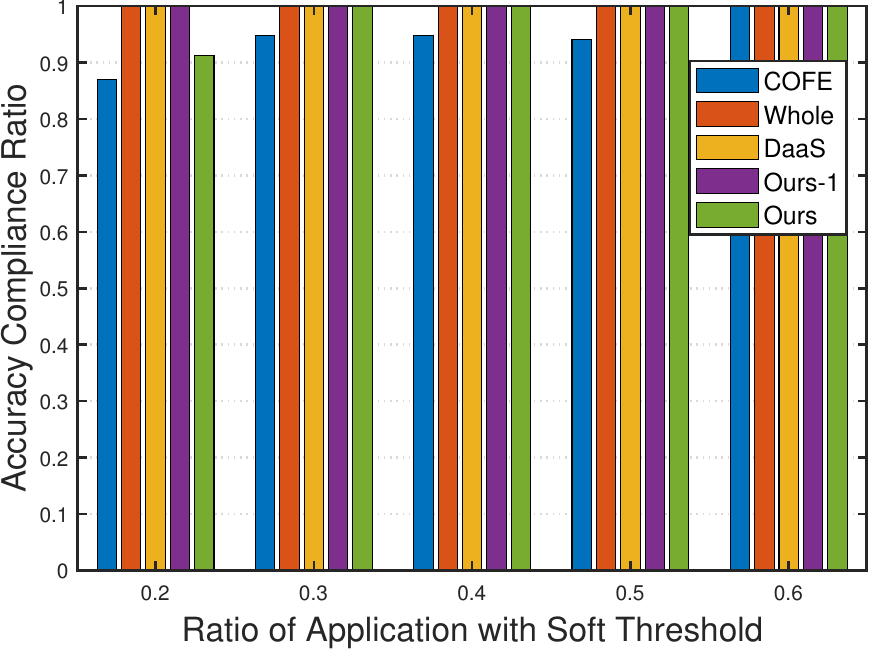}\captionsetup{font={sf,small}}}
     \vspace{0.1cm} 
    \subfloat[]{\includegraphics[width=0.32\linewidth]{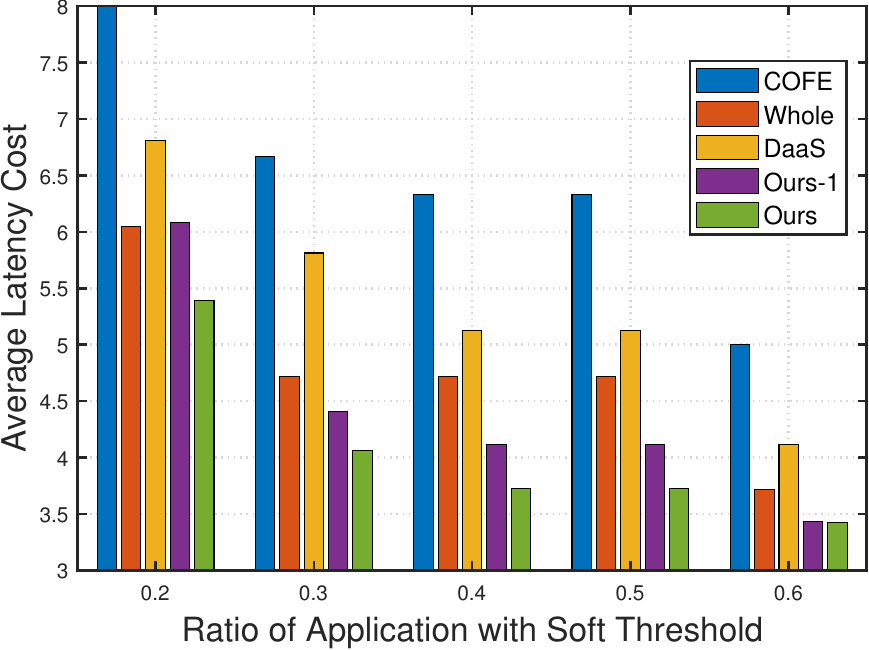}\captionsetup{font={sf,small}}}
\hspace{0.01\linewidth} 
    \subfloat[]{\includegraphics[width=0.32\linewidth]{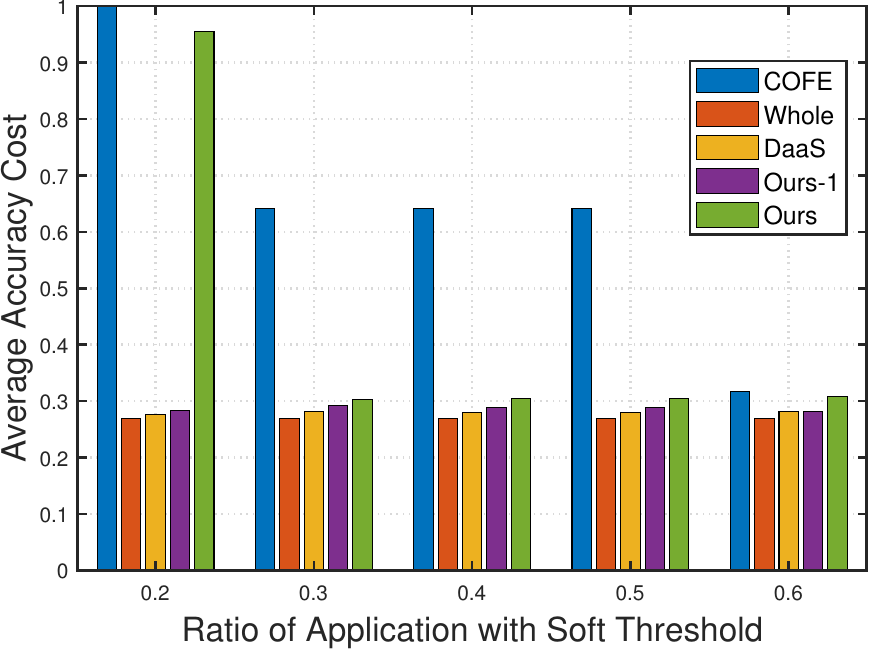}\captionsetup{font={sf,small}}}
\hspace{0.01\linewidth} 
    \subfloat[]{\includegraphics[width=0.32\linewidth]{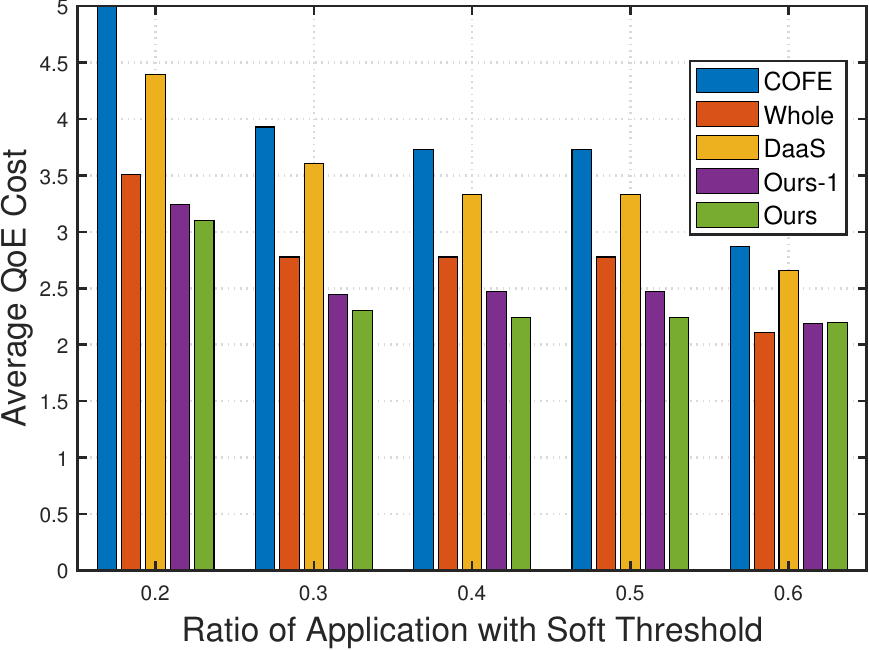}\captionsetup{font={sf,small}}}

    \caption{Impact of the ratio of application with soft threshold. (a) Average completion time versus the ratio of application with soft threshold. (b) Deadline compliance ratio versus the ratio of application with soft threshold. (c) Accuracy compliance ratio versus the ratio of application with soft threshold. (d) Average latency cost versus the ratio of application with soft threshold. (e) Average accuracy cost versus the ratio of application with soft threshold. (f) Average QoE cost versus the ratio of application with soft threshold.}
\end{figure*}
\begin{figure}
    \centering
     \setstretch{0.0}
\renewcommand{\sfdefault}{ptm}
    \subfloat[]{\includegraphics[width=0.85\linewidth]{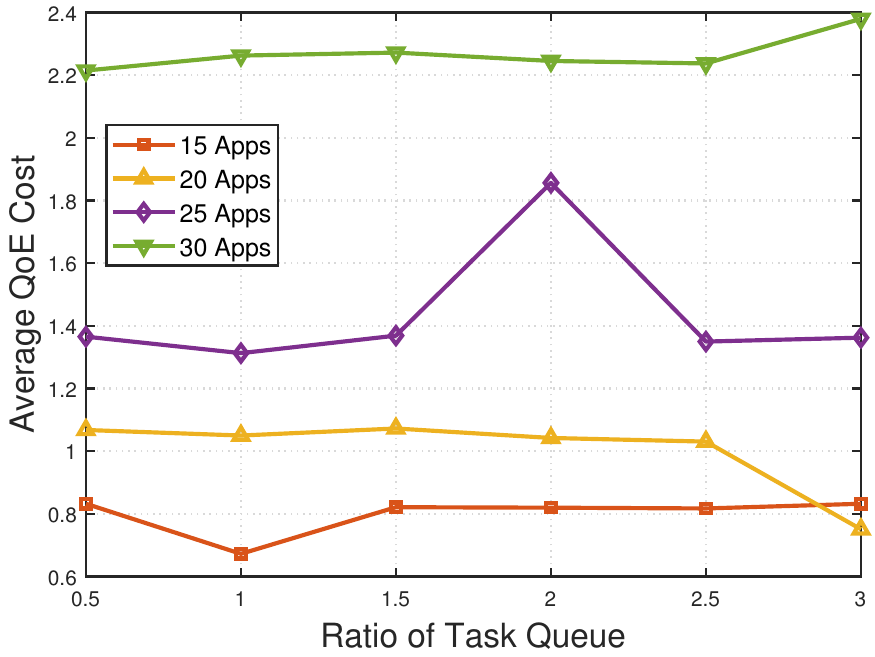}\captionsetup{font={sf,footnotesize}}}

    \subfloat[]{\includegraphics[width=0.85\linewidth]{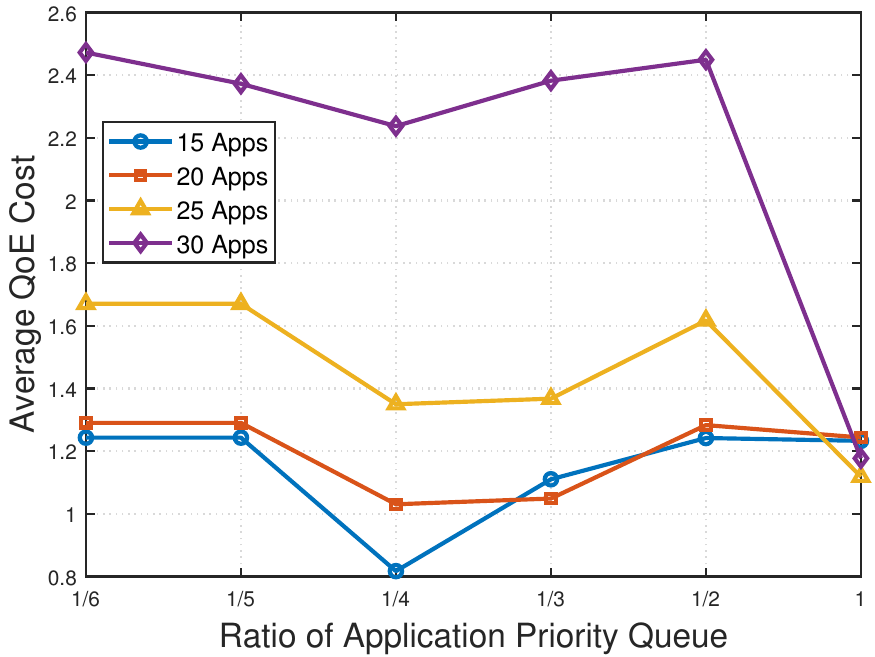}\captionsetup{font={sf,small}}}
    
    \caption{Impact of parameters of our algorithm. (a)Average QoE cost versus the ratio of task queue.(b) Average QoE cost versus the ratio of application priority queue.}
\end{figure}
\subsubsection{Impact of the Configuration of Network}
We set the number of App Node as 30 and adjust the number of node of the mesh network to test algorithm performance using the six metrics as shown in Fig. 6. As the number of nodes increases, the average completion time exhibits declining trends. However, due to dependencies among tasks and limitations of application deadline, both the deadline compliance ratio and average latency cost decrease to some extend and stabilize with the raising number of nodes. Conversely, with the increasing number of nodes, the transmission links increase, leading to an improvement in the accuracy of applications as shown in Fig. 6(c)(e). Consequently, the average of QoE cost predominantly relies on the deadline constraint, exhibiting a similar trend to the average latency cost.

It can be shown in the Fig. 6 that when the number of nodes is small (e.g. 40), the all metrics of COFE exhibit the worst performance. However, as the number of nodes increases,the performance of COFE inproves, achieving the slightly higher deadline compliance ratio, lower average latency cost and similar average QoE cost with ours. This can be attributed to COFE's task assignment strategy, which assigns tasks only when predecessor tasks are completed, without prioritizing tasks based on deadlines in advance. The performances of Whole wave depending on the randomness and limitation of capacity of one node. In contrast, our algorithm consistently achieves the lowest average QoE cost compared to other baselines especially when the number of nodes is small. It can be observed in Fig. 6(f), as the number of nodes increases (e.g., 60), leading to an increase in available resources, the differences in algorithm performance decrease. 
\subsubsection{Impact of the Ratio of Application with Soft Threshold}
In order to evaluate the performance of our proposed algorithm, we set the number of App Node to 30 and the number of node to 40, respectively. We then adjust the ratio of the application with soft threshold in the mesh network and observe the performance of the six metrics as shown in Fig. 7. Notably, as the ratio of applications with a soft threshold increases, the average completion time and deadline compliance ratio stabilize. This stabilization is attributed to the limited computing resources and transmission links, constrained by the number of nodes. Conversely, when the ratio is reduced, more applications can tolerate slight degradation, resulting in an increase in accuracy compliance ratio, and a decrease in average latency cost and average accuracy cost. Consequently, the average QoE cost exhibits a declining trend.

As a ratio is 0.6 for applications with a soft threshold, the average QoE cost of Whole is slightly lower than that of Ours-1 and Ours. This is attributed to the lower average accuracy cost playing a more important role than the average latency cost. As the ratio of application of soft threshold increases, indicating a higher demand for guarantee for deadline and data accuracy, Whole struggles to fully leverage computing resource, leading to higher latency cost and subsequently, higher QoE cost. COFE can not guarantee any applications with hard deadline threshold. With the ratio of application with soft threshold increasing, the differences of performance from DaaS and COFE to Ours reduce in terms of average QoE cost. Compared to Ours-1, Ours leans towards prioritizing latency over accuracy, achieving lower average latency costs while incurring higher average accuracy costs. Ours outperforms Ours-1 when the ratio of applications with a soft threshold reduces.
\subsubsection{Impact of the Ratio of Appication Priority Queue and Task Queue}
To determine the optimal parameter settings for our algorithm, we conducted experiments with the ratio of application priority queue $k$ and the ratio of task queue $o$. As shown in Fig. 8(a), our algorithm achieves a lower average QoE cost when the ratio of task queue is set to 1, indicating that the size of the task queue equals the number of task of its application. However, as the ratio increases, the average QoE cost exhibits fluctuations. This is because when the task size is large, our algorithm may assign all tasks of an application, leading to varying performance outcomes. To enhance the robustness of our algorithm, we set the ratio of queue to 1, which reduces the round of task scheduling campared to smaller values and ensures consistent good performance.

Fig. 8(b) illustrates the impact of the ratio of application priority queue. It can be observed that our algorithm performs better when all applications or 1/4 of the applications are ready for assignment. Additionally, in cases where the number of applications is relatively small (e.g., 15, 20), the average QoE cost is lower when the ratio is set to 1/4 compared to a ratio of 1. The reason is that a balanced assignment scheme tends to result in lower average QoE when a large number of applications requesting to process. Conversely, with a smaller number of applications, assigning more tasks of an application in a scheduling round yields a better performance. However, as the ratio decreases (e.g., 1/6, 1/5), it will aggravate to assign the tasks from the same application according to application priority, which may lead to performance deterioration.
\section{Conclusions}
In this paper, we propose a QoE cost model to evaluate the completion of dependent tasks under multi-dimensional QoS constraints. This model is constructed by creating performance degradation functions for diverse dimensions and sensitivities of QoS, and these functions are amalgamated into the unified QoE cost model that considers application-specific QoS preferences. Futhermore, in order to minimize the overall QoE cost in the distributed networks, we present a HMTSA, which employs hierarchical multiple queues to determine the amount of task of all applications and priorities of these tasks and choose the optimal node to execute in each round of scheduling. Extensive simulation results demonstrate the effectiveness of our algorithm in scheduling tasks across networks, particularly under varying ratios of QoS sensitivity, diverse application counts, and varying numbers of computing nodes.

\newpage
%
%
%
%
%

\vfill

\end{document}